\documentclass
	[
		10 pt
	]
	{article}
\usepackage
	[
		a4paper,
		left = 2.500 cm,
		right = 2.500 cm,
		top = 2.500 cm,
		bottom = 3.000 cm
	]
	{geometry}
\usepackage{import}
\usepackage{tikz, pgf, pgfplots}
\usetikzlibrary
	{
		3d,
		angles,
		arrows,											%	Customizing arrows
		arrows.meta,
		automata,										%	Import library for drawing automata 
		backgrounds,
		bending,
		calc,
		chains,
		datavisualization.formats.functions,			%	Plotting functions
		decorations.pathmorphing,
		decorations.pathreplacing,
		decorations.text,
		fadings,
		fit,
		graphs,
		graphs.standard,
		matrix,
		patterns,
		positioning,									%	Positioning nodes
		quantikz2,
		quotes,
		shadings,
		shadows,
		shadows.blur,
		shapes,
		shapes.geometric,
		trees
	}
\tikzset{
	every state/.style = {
		circle, draw, semithick,
		minimum size = 10.000 mm,			%	uniform circles (remove for text-sized circles)
		fill = WordAquaAccent2Lighter40,
	},
	initial text = {},
	every initial by arrow/.style = ->,
	double distance = 2.000 pt,
	every edge/.style = {
		->, draw,
		> = { Latex [round] },
		shorten >= 1.000 pt,
		auto,
		semithick
	}
}
\pgfplotsset{compat = newest}
\usepgflibrary
	{
		shapes.misc
	}
\usepgfplotslibrary
	{
		dateplot
	}
\usepackage{tikzpeople}
\usepackage{fontawesome6}
\usepackage{smartdiagram}
\usepackage{amsmath}
\usepackage{amssymb}
\usepackage{amsthm}
\allowdisplaybreaks
\usepackage{refcount}
\usepackage[bb = boondox]{mathalfa}
\usepackage{dsfont}
\usepackage{newtxmath}
\usepackage{mathtools}
\usepackage{multicol}
\usepackage{physics2}
\usephysicsmodule{ab.legacy}
\usephysicsmodule{braket}
\usephysicsmodule{diagmat}
\usephysicsmodule{nabla.legacy}
\usepackage{empheq}
\usepackage[ compat = newest ]{yquant}
\usepackage{sectsty}
\allsectionsfont{\color{WordBlue}}
\usepackage[x11names]{xcolor}
\usepackage{varwidth}
\usepackage{graphicx}
\usepackage{wrapfig}
\usepackage{subcaption}
\usepackage{xurl}                                                                                  %
\usepackage{hyperref}
\hypersetup
	{
		colorlinks,
		linkcolor = {RedPurple},
		citecolor = {WordBlueDarker25},
		urlcolor = {WordAquaDarker50}
	}
\usepackage{tabularray}
\usepackage{booktabs}
\usepackage{colortbl}
\usepackage{float}
\usepackage{nicematrix}
\usepackage{cellspace}
\usepackage{makecell}
\usepackage{multirow}
\usepackage{caption}
\usepackage[ linesnumbered, tworuled, vlined ] {algorithm2e}
\usepackage{appendix}
\usepackage{enumitem}
\usepackage{siunitx}
\usepackage{xintexpr}
\usepackage{spreadtab}
\usepackage{framed}
\usepackage{svg}
\usepackage{lipsum}
\usepackage{listings}
\definecolor{DarkIndianRed}{HTML}{8C3232}
\definecolor{MyLightRed}{RGB}{244, 213, 245}
\definecolor{Purple}{HTML}{911146}
\definecolor{PurpleDark}{RGB}{102, 0, 102}
\definecolor{PurpleDS}{HTML}{9650C8}
\definecolor{RedDarkLight}{HTML}{EA005F}
\definecolor{RedDarkLightest}{HTML}{FF0088}
\definecolor{RedPurple}{HTML}{AA007F}
\definecolor{RedPurpleDS}{HTML}{C83278}
\definecolor{WordPinkAccent1Darker25}{HTML}{B3186D}
\definecolor{WordPinkAccent1Darker50}{HTML}{781049}
\definecolor{WordPinkAccent1Lighter40}{HTML}{EE80BC}
\definecolor{WordPinkAccent1Lighter60}{HTML}{F3AAD2}
\definecolor{WordPinkAccent1Lighter80}{HTML}{F9D4E8}
\definecolor{WordRed}{RGB}{255, 0, 102}
\definecolor{WordRedAccent5Lighter60}{HTML}{F5B5A7}
\definecolor{WordRedAccent5Darker25}{HTML}{B23214}
\definecolor{MagentaDark}{RGB}{106, 65, 152}
\definecolor{MagentaDS}{HTML}{C85096}
\definecolor{MagentaLight}{RGB}{128, 100, 162}
\definecolor{MagentaLighter}{RGB}{161, 106, 221}
\definecolor{MagentaVeryDark}{RGB}{97, 75, 128}
\definecolor{MagentaVeryLight}{RGB}{178, 162, 201}
\definecolor{OrangeDS}{HTML}{E67832}
\definecolor{WordOrangeAccent2Lighter60}{HTML}{FCD3A4}
\definecolor{WordOrangeAccent4Lighter60}{HTML}{F7C5A1}
\definecolor{GreenDark}{HTML}{225522}
\definecolor{GreenDS}{HTML}{82C896}
\definecolor{GreenLighter1}{HTML}{00B383}
\definecolor{GreenLighter2}{HTML}{00AA7F}
\definecolor{GreenLightest}{HTML}{00FFA0}
\definecolor{GreenTeal}{HTML}{008080}
\definecolor{WordLightGreen}{RGB}{140, 214, 192}
\definecolor{WordGreen}{RGB}{0, 176, 80}
\definecolor{BlueDS}{HTML}{3278C8}
\definecolor{BlueVeryDark}{HTML}{222255}
\definecolor{MyBlue}{RGB}{0, 64, 128}
\definecolor{MyDarkBlue}{RGB}{0, 51, 102}
\definecolor{MyVeryLightBlue}{RGB}{211, 245, 247}
\definecolor{WordBlue}{RGB}{19, 65, 99}
\definecolor{WordBlueDark}{RGB}{46, 116, 181}
\definecolor{WordBlueDarker}{RGB}{31, 78, 121}
\definecolor{WordBlueDarker25}{RGB}{54, 96, 146}
\definecolor{WordBlueDarker50}{RGB}{36, 64, 98}
\definecolor{WordBlueDarkest}{RGB}{0, 32, 96}
\definecolor{WordBlueLight}{RGB}{0, 112, 192}
\definecolor{WordBlueVeryLight}{HTML}{00B0F0}
\definecolor{WordIceBlue}{RGB}{223, 227, 229}
\definecolor{WordAquaAccent1Darker25}{HTML}{276E8B}
\definecolor{WordAquaAccent1Darker50}{HTML}{1A495D}
\definecolor{WordAquaAccent1Lighter40}{HTML}{7FC0DB}
\definecolor{WordAquaAccent1Lighter60}{HTML}{A9D5E7}
\definecolor{WordAquaAccent1Lighter80}{HTML}{D4EAF3}
\definecolor{WordAquaAccent2Darker25}{HTML}{398E98}
\definecolor{WordAquaAccent2Darker50}{HTML}{265F65}
\definecolor{WordAquaAccent2Lighter40}{HTML}{9AD3D9}
\definecolor{WordAquaAccent2Lighter60}{HTML}{BCE1E5}
\definecolor{WordAquaAccent2Lighter80}{HTML}{DDF0F2}
\definecolor{WordAquaDarker25}{HTML}{31869B}
\definecolor{WordAquaDarker50}{HTML}{215967}
\definecolor{WordAquaLighter40}{HTML}{92CDDC}
\definecolor{WordAquaLighter60}{HTML}{B7DEE8}
\definecolor{WordAquaLighter80}{HTML}{DAEEF3}
\definecolor{CyanDS}{HTML}{50B4D2}
\definecolor{WordDarkerTeal}{RGB}{48, 82, 80}
\definecolor{WordDarkTeal}{RGB}{72, 123, 119}
\definecolor{WordDarkTealLighter80}{RGB}{207, 223, 234}
\definecolor{WordLightTeal}{RGB}{160, 199, 197}
\definecolor{WordVeryLightTeal}{RGB}{223, 236, 235}
\definecolor{WordTurquoiseLighter80}{RGB}{209, 238, 249}
\definecolor{Brown}{HTML}{666633}
\definecolor{WordGoldAccent1Darker25}{HTML}{C49A00}
\definecolor{WordGoldAccent1Lighter40}{HTML}{FFDF6A}
\definecolor{YellowDS}{HTML}{F0D228}
\definecolor{LavenderBlush}{RGB}{255, 240, 245}
\definecolor{MediumTurquoise}{RGB}{72, 209, 204}
\definecolor{PowderBlue}{RGB}{176, 224, 230}
\definecolor{SkyBlue}{RGB}{135, 206, 235}
\definecolor{Azure2}{RGB}{224, 238, 238}
\definecolor{Azure3}{RGB}{193, 205, 205}
\definecolor{CadetBlue4}{RGB}{83, 134, 139}
\definecolor{DarkSeaGreen1}{RGB}{193, 255, 193}
\definecolor{DeepPink4}{RGB}{139, 10, 80}
\definecolor{Honeydew2}{RGB}{224, 238, 224}
\definecolor{LightSkyBlue1}{RGB}{176, 226, 255}
\definecolor{LightSkyBlue3}{RGB}{141, 182, 205}
\definecolor{LightSkyBlue4}{RGB}{96, 123, 139}
\definecolor{LightSteelBlue1}{RGB}{202, 225, 255}
\definecolor{LightSteelBlue4}{RGB}{110, 123, 139}
\definecolor{MediumPurple1}{RGB}{171, 130, 255}
\definecolor{PaleTurquoise3}{RGB}{150, 205, 205}
\definecolor{PaleVioletRed3}{RGB}{205, 104, 137}
\definecolor{Purple1}{RGB}{155, 48, 255}
\definecolor{SeaGreen1}{RGB}{84, 255, 159}
\definecolor{SeaGreen2}{RGB}{78, 238, 148}
\definecolor{SeaGreen3}{RGB}{67, 205, 128}
\definecolor{SkyBlue1}{HTML}{87CEFF}
\definecolor{SkyBlue4}{RGB}{74, 112, 139}
\definecolor{SteelBlue1}{RGB}{99, 184, 255}
\definecolor{Thistle3}{RGB}{205, 181, 205}
\definecolor{Turquoise4}{RGB}{0, 134, 139}
\definecolor{VioletRed1}{RGB}{255, 62, 150}
\definecolor{VioletRed2}{RGB}{208, 32, 144}
\definecolor{VioletRed3}{RGB}{199, 21, 133}
\definecolor{VioletRed4}{RGB}{139, 10, 80}
\usepackage
	[
		most
	]
	{tcolorbox}
\tcbset{
	MyInfoPanel/.style = {
		enhanced, sharp corners,
		colback = Seashell1!50,
		boxrule = 0.400 mm,
		frame hidden,
		overlay = {
			\draw [ Seashell3, line width = 0.400 mm ]
			([ xshift = - 10.000 mm ]frame.north east)
			-- (frame.north east)
			-- ([ yshift = - 10.000 mm ]frame.north east);
			\draw [ Seashell3, line width = 0.400 mm ]
			([ xshift = 10.000 mm ]frame.south west)
			-- (frame.south west)
			-- ([ yshift = 10.000 mm ]frame.south west);
		},
		left = 1.000 mm, right = 1.000 mm, top = 1.000 mm, bottom = 1.000 mm,
		width=\linewidth
	}
}
\tcbset{
	MyFigurePanel/.style = {
		enhanced, sharp corners,
		colback = WordIceBlue!25,
		boxrule = 0.400 mm,
		frame hidden,
		overlay = {
			\draw [ WordIceBlue, line width = 0.400 mm ]
			([ xshift = 10.000 mm ]frame.north west)
			-- (frame.north west)
			-- ([ yshift = - 10.000 mm ]frame.north west);
			\draw [ WordIceBlue, line width = 0.400 mm ]
			([ xshift = - 10.000 mm ]frame.south east)
			-- (frame.south east)
			-- ([ yshift = 10.000 mm ]frame.south east);
		},
		left = 1.000 mm, right = 1.000 mm, top = 1.000 mm, bottom = 1.000 mm,
		width=\linewidth
	}
}
\tcbset{
	MyInfoBoxStyle/.style = {
		breakable, enhanced jigsaw,
		colbacktitle = CyanDS,
		coltitle = black,
		colback = WordAquaAccent1Lighter80!40,
		colframe = white,
		arc = 4.000 mm,
		boxrule = 0.000 pt,
		center title, fonttitle = \bfseries,
		varwidth boxed title
	}
}
\newtcolorbox{MyInfoBox}[1][Synopsis]{MyInfoBoxStyle, title = {#1}}
\newtcolorbox{MyGeneralEnvironment} [ 1 ] [ ]
	{
		breakable,
		enhanced,
		enhanced jigsaw,
		attach boxed title to top left = { xshift = -5.000 mm, yshift = 0.000 mm },
		boxed title style = { boxrule = 0.000 pt, sharp corners = all },
		colbacktitle = PaleVioletRed3!50,
		coltitle = black,
		fonttitle = \bfseries,
		varwidth boxed title,
		colback = WordPinkAccent1Lighter80!12,
		colframe = WordPinkAccent1Darker50,
		sharp corners = all,
		toprule = 0.000 mm,
		bottomrule = 0.500 mm,
		leftrule = 0.500 mm,
		rightrule = 0.000 mm,
		title = {#1},
	}
\newcounter{MyCorollary}[section]
\renewcommand{\theMyCorollary}{\thesection.\arabic{MyCorollary}}
\newtcolorbox{corollary} [ 1 ] [ ]
	{
		breakable,
		enhanced,
		enhanced jigsaw,
		skin = enhanced,
		attach boxed title to top left = { xshift = -5.000 mm, yshift = 0.000 mm },
		boxed title style = { boxrule = 0.000 pt, sharp corners = all },
		colbacktitle = RedPurple!90!black,
		coltitle = white,
		fonttitle = \bfseries,
		varwidth boxed title,
		colback = RedPurple!03,
		colframe = RedPurple,
		sharp corners = all,
		toprule = 0.000 mm,
		bottomrule = 0.500 mm,
		leftrule = 0.500 mm,
		rightrule = 0.000 mm,
		code = { \refstepcounter{MyCorollary} },
		title = {Corollary~\theMyCorollary:\if\relax\detokenize{#1}\relax\else~#1\fi},
	}
\newtcolorbox{corollary*} [ 1 ] [ ]
	{
		breakable,
		enhanced,
		enhanced jigsaw,
		skin = enhanced,
		attach boxed title to top left = { xshift = -5.000 mm, yshift = 0.000 mm },
		boxed title style = { boxrule = 0.000 pt, sharp corners = all },
		colbacktitle = RedPurple!90!black,
		coltitle = white,
		fonttitle = \bfseries,
		varwidth boxed title,
		colback = RedPurple!03,
		colframe = RedPurple,
		sharp corners = all,
		toprule = 0.000 mm,
		bottomrule = 0.500 mm,
		leftrule = 0.500 mm,
		rightrule = 0.000 mm,
		title = {Corollary\if\relax\detokenize{#1}\relax\else~#1\fi},
	}
\newcounter{MyDefinition}[section]
\renewcommand{\theMyDefinition}{%
	\number\numexpr\value{section}-1\relax.%
\ifnum\value{MyDefinition}<10 0\else\ifnum\value{MyDefinition}<100 \fi\fi\arabic{MyDefinition}}
\newtcolorbox{definition} [ 1 ] [ ]
	{
		breakable,
		enhanced,
		enhanced jigsaw,
		skin = enhanced,
		attach boxed title to top left = { xshift = -5.000 mm, yshift = 0.000 mm },
		boxed title style = { boxrule = 0.000 pt, sharp corners = all },
		colbacktitle = SkyBlue!70,
		coltitle = SkyBlue!30!black,
		fonttitle = \bfseries,
		varwidth boxed title,
		colback = SkyBlue!15,
		colframe = SkyBlue,
		sharp corners = all,
		toprule = 0.000 mm,
		bottomrule = 0.500 mm,
		leftrule = 0.500 mm,
		rightrule = 0.000 mm,
		code = { \refstepcounter{MyDefinition} },
		title = {Definition~\theMyDefinition:\if\relax\detokenize{#1}\relax\else~#1\fi},
	}
\newtcolorbox{definition*} [ 1 ] [ ]
	{
		breakable,
		enhanced,
		enhanced jigsaw,
		skin = enhanced,
		attach boxed title to top left = { xshift = -5.000 mm, yshift = 0.000 mm },
		boxed title style = { boxrule = 0.000 pt, sharp corners = all },
		colbacktitle = SkyBlue!70,
		coltitle = SkyBlue!30!black,
		fonttitle = \bfseries,
		varwidth boxed title,
		colback = SkyBlue!15,
		colframe = SkyBlue,
		sharp corners = all,
		toprule = 0.000 mm,
		bottomrule = 0.500 mm,
		leftrule = 0.500 mm,
		rightrule = 0.000 mm,
		title = {Definition\if\relax\detokenize{#1}\relax\else~#1\fi},
	}
\newcounter{MyExample}[section]
\renewcommand{\theMyExample}{%
	\number\numexpr\value{section}-1\relax.%
\ifnum\value{MyExample}<10 0\else\ifnum\value{MyExample}<100 \fi\fi\arabic{MyExample}}
\newtcolorbox{example} [ 1 ] [ ]
	{
		breakable,
		enhanced,
		enhanced jigsaw,
		skin = enhanced,
		attach boxed title to top left = { xshift = -5.000 mm, yshift = 0.000 mm },
		boxed title style = { boxrule = 0.000 pt, sharp corners = all },
		colbacktitle = WordAquaAccent1Darker25,
		coltitle = white,
		fonttitle = \bfseries,
		varwidth boxed title,
		colback = WordAquaAccent1Lighter80!40,
		coltext = WordAquaAccent1Darker25!50!black,
		colframe = WordAquaAccent1Darker25,
		sharp corners = all,
		toprule = 0.000 mm,
		bottomrule = 0.500 mm,
		leftrule = 0.500 mm,
		rightrule = 0.000 mm,
		code = { \refstepcounter{MyExample} },
		title = {Example~\theMyExample:\if\relax\detokenize{#1}\relax\else~#1\fi},
	}
\newtcolorbox{example*} [ 1 ] [ ]
	{
		breakable,
		enhanced,
		enhanced jigsaw,
		skin = enhanced,
		attach boxed title to top left = { xshift = -5.000 mm, yshift = 0.000 mm },
		boxed title style = { boxrule = 0.000 pt, sharp corners = all },
		colbacktitle = WordAquaAccent1Darker25,
		coltitle = white,
		fonttitle = \bfseries,
		varwidth boxed title,
		colback = WordAquaAccent1Lighter80!25,
		colframe = WordAquaAccent1Darker25,
		sharp corners = all,
		toprule = 0.000 mm,
		bottomrule = 0.500 mm,
		leftrule = 0.500 mm,
		rightrule = 0.000 mm,
		title = {Example\if\relax\detokenize{#1}\relax\else~#1\fi},
	}
\newcounter{MyLemma}[section]
\renewcommand{\theMyLemma}{\thesection.\arabic{MyLemma}}
\newtcolorbox{lemma} [ 1 ] [ ]
	{
		breakable,
		enhanced,
		enhanced jigsaw,
		skin = enhanced,
		attach boxed title to top left = { xshift = -5.000 mm, yshift = 0.000 mm },
		boxed title style = { boxrule = 0.000 pt, sharp corners = all },
		colbacktitle = PaleVioletRed3!50,
		coltitle = black,
		fonttitle = \bfseries,
		varwidth boxed title,
		colback = WordPinkAccent1Lighter80!12,
		colframe = WordPinkAccent1Darker50,
		sharp corners = all,
		toprule = 0.000 mm,
		bottomrule = 0.500 mm,
		leftrule = 0.500 mm,
		rightrule = 0.000 mm,
		code = { \refstepcounter{MyLemma} },
		title = {Lemma~\the\theMyLemma:\if\relax\detokenize{#1}\relax\else~#1\fi},
	}
\newtcolorbox{lemma*} [ 1 ] [ ]
	{
		breakable,
		enhanced,
		enhanced jigsaw,
		skin = enhanced,
		attach boxed title to top left = { xshift = -5.000 mm, yshift = 0.000 mm },
		boxed title style = { boxrule = 0.000 pt, sharp corners = all },
		colbacktitle = PaleVioletRed3!50,
		coltitle = black,
		fonttitle = \bfseries,
		varwidth boxed title,
		colback = WordPinkAccent1Lighter80!12,
		colframe = WordPinkAccent1Darker50,
		sharp corners = all,
		toprule = 0.000 mm,
		bottomrule = 0.500 mm,
		leftrule = 0.500 mm,
		rightrule = 0.000 mm,
		title = {Lemma\if\relax\detokenize{#1}\relax\else~#1\fi},
	}
\newcounter{MyProposition}[section]
\renewcommand{\theMyProposition}{\thesection.\arabic{MyProposition}}
\newtcolorbox{proposition} [ 1 ] [ ]
	{
		breakable,
		enhanced,
		enhanced jigsaw,
		skin = enhanced,
		attach boxed title to top left = { xshift = -5.000 mm, yshift = 0.000 mm },
		boxed title style = { boxrule = 0.000 pt, sharp corners = all },
		colbacktitle = cyan7!50,
		coltitle = black,
		fonttitle = \bfseries,
		varwidth boxed title,
		colback = cyan9!25,
		colframe = cyan5,
		sharp corners = all,
		toprule = 0.000 mm,
		bottomrule = 0.500 mm,
		leftrule = 0.500 mm,
		rightrule = 0.000 mm,
		code = { \refstepcounter{MyProposition} },
		title = {Proposition~\theMyProposition:\if\relax\detokenize{#1}\relax\else~#1\fi},
	}
\newtcolorbox{proposition*} [ 1 ] [ ]
	{
		breakable,
		enhanced,
		enhanced jigsaw,
		skin = enhanced,
		attach boxed title to top left = { xshift = -5.000 mm, yshift = 0.000 mm },
		boxed title style = { boxrule = 0.000 pt, sharp corners = all },
		colbacktitle = cyan7!50,
		coltitle = black,
		fonttitle = \bfseries,
		varwidth boxed title,
		colback = cyan9!25,
		colframe = cyan5,
		sharp corners = all,
		toprule = 0.000 mm,
		bottomrule = 0.500 mm,
		leftrule = 0.500 mm,
		rightrule = 0.000 mm,
		title = {Proposition\if\relax\detokenize{#1}\relax\else~#1\fi},
	}
\newcounter{MyTheorem}[section]
\renewcommand{\theMyTheorem}{%
	\number\numexpr\value{section}-1\relax.%
	\ifnum\value{MyTheorem}<10 0\else\ifnum\value{MyTheorem}<100 \fi\fi\arabic{MyTheorem}
}
\newtcolorbox{theorem} [ 1 ] [ ]
	{
		breakable,
		enhanced,
		enhanced jigsaw,
		skin = enhanced,
		attach boxed title to top left = { xshift = -5.000 mm, yshift = 0.000 mm },
		boxed title style = { boxrule = 0.000 pt, sharp corners = all },
		colbacktitle = WordAquaAccent2Darker25,
		coltitle = white,
		fonttitle = \bfseries,
		varwidth boxed title,
		colback = WordAquaAccent2Lighter80,
		colframe = WordAquaAccent2Darker25,
		sharp corners = all,
		toprule = 0.000 mm,
		bottomrule = 0.500 mm,
		leftrule = 0.500 mm,
		rightrule = 0.000 mm,
		code = { \refstepcounter{MyTheorem} },
		title = {Theorem~\theMyTheorem:\if\relax\detokenize{#1}\relax\else~#1\fi},
	}
\newtcolorbox{theorem*} [ 1 ] [ ]
	{
		breakable,
		enhanced,
		enhanced jigsaw,
		skin = enhanced,
		attach boxed title to top left = { xshift = -5.000 mm, yshift = 0.000 mm },
		boxed title style = { boxrule = 0.000 pt, sharp corners = all },
		colbacktitle = WordAquaAccent2Darker25,
		coltitle = white,
		fonttitle = \bfseries,
		varwidth boxed title,
		colback = WordAquaAccent2Lighter80,
		colframe = WordAquaAccent2Darker25,
		sharp corners = all,
		toprule = 0.000 mm,
		bottomrule = 0.500 mm,
		leftrule = 0.500 mm,
		rightrule = 0.000 mm,
		title = {Theorem\if\relax\detokenize{#1}\relax\else~#1\fi},
	}
\newcounter{mathseed}
\pgfmathsetseed{\arabic{mathseed}}
\pgfdeclarelayer{background}
\pgfsetlayers{background, main}
\pgfdeclaredecoration{irregular fractal line}{init}
{
	\state{init}[width=\pgfdecoratedinputsegmentremainingdistance]
	{
		\pgfpathlineto{\pgfpoint{random*\pgfdecoratedinputsegmentremainingdistance}{(random*\pgfdecorationsegmentamplitude-0.02)*\pgfdecoratedinputsegmentremainingdistance}}
		\pgfpathlineto{\pgfpoint{\pgfdecoratedinputsegmentremainingdistance}{0pt}}
	}
}
\def\tornpaper#1{%
	\ifthenelse{\isodd{\value{mathseed}}}
	{%
		\tikz
		{
			\node[inner sep = 1em] (A) {#1};		%	Draw the text of the node
			\begin{pgfonlayer}{background}			%	Draw the shape behind
				\fill[paper]						%	Recursively decorate the bottom border
				\pgfextra{\pgfmathsetseed{\arabic{mathseed}}\addtocounter{mathseed}{1}}%
				{decorate[irregular cloudy border]{decorate{decorate{decorate{decorate[ragged border]{
										(A.north west) -- (A.north east)
				}}}}}}
				-- (A.south east)
				\pgfextra{\pgfmathsetseed{\arabic{mathseed}}}%
				{decorate[irregular spiky border]{decorate{decorate{decorate{decorate[ragged border]{
										-- (A.south west)
				}}}}}}
				-- (A.north west);
			\end{pgfonlayer}
		}
	}
	{%
		\tikz{
			\node[inner sep=1em] (A) {#1};  % Draw the text of the node
			\begin{pgfonlayer}{background}  % Draw the shape behind
				\fill[paper] % recursively decorate the bottom border
				\pgfextra{\pgfmathsetseed{\arabic{mathseed}}\addtocounter{mathseed}{1}}%
				{decorate[irregular spiky border]{decorate{decorate{decorate{decorate[ragged border]{
										(A.north east) -- (A.north west)
				}}}}}}
				-- (A.south west)
				\pgfextra{\pgfmathsetseed{\arabic{mathseed}}}%
				{decorate[irregular cloudy border]{decorate{decorate{decorate{decorate[ragged border]{
										-- (A.south east)
				}}}}}}
				-- (A.north east);
		\end{pgfonlayer}}
	}
}
\usepackage [ scale = 2 ] {ccicons}
\usepackage{changepage}
\usepackage{standalone}
\title
	{
		Where Does a Variational Quantum Classifier Lose Accuracy? Angle Encoding, Readout and the Fidelity-Kernel Reference in Cyberattack Detection
	}
\usepackage{orcidlink}
\newcommand{\orcidicon}[1]{\href{https://orcid.org/#1}{\includegraphics[height=\fontcharht\font`\B]{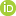}}}
\author
	{
		Angelos Thomos$^1$\orcidicon{0009-0003-1340-8515}
		and
		Theodore Andronikos$^1$\orcidicon{0000-0002-3741-1271}
		\\[10 pt]
		$^1$
		Department of Informatics, Ionian University, \\
		7 Tsirigoti Square, 49100 Corfu, Greece; \\
		\{p19thom, andronikos\}@ionio.gr
	}
\begin{document}
%                                                                                                  %
%--------------------------------------------------------------------------------------------------%
%--------------------------------------------------------------------------------------------------%
%--------------------------------------------------------------------------------------------------%

%--------------------------------------------------------------------------------------------------%
%                                                                                                  %
%			G e n e r a t e   t h e   T i t l e                                                    %
%                                                                                                  %
%--------------------------------------------------------------------------------------------------%
%                                                                                                  %
\maketitle
%                                                                                                  %
%--------------------------------------------------------------------------------------------------%

%--------------------------------------------------------------------------------------------------%
%                                                                                                  %
%			A b s t r a c t                                                                        %
%                                                                                                  %
%--------------------------------------------------------------------------------------------------%
%                                                                                                  %
\begin{abstract}
	When a variational quantum classifier performs poorly at detecting network attacks, it is hard to tell which part is to blame: the encoding, the trained circuit or the readout, which turns measurements into a class. In this work, we build upon a recent amplitude-encoded classifier, and we try to separate their contributions to the NSL-KDD benchmark in noiseless simulation. We compare angle and amplitude encoding, vary one component at a time, and measure each circuit against a classical yardstick: a support-vector machine on the encoded states, whose accuracy estimates what a well-trained circuit can reach. Most of the shortfall lies in the readout. Reading one qubit, or the parity of all qubits, leaves several percentage points unused, while a readout trained together with the circuit recovers most of them from the same measurements. This also explains why some classifiers stop improving as further layers are added. The encodings suit different needs. Angle encoding gives the more accurate detector on traffic like the training data and, from eight features up, shallower circuits; amplitude encoding needs far fewer qubits and catches more attacks of types never seen in training. Finally, we report that classical models, given the same inputs, are about as accurate as the best circuits, thus, registering no quantum advantage.
	\\[12 pt]
%                                                                                                  %
%--------------------------------------------------------------------------------------------------%
%                                                                                                  %
%			K e y w o r d s                                                                        %
%                                                                                                  %
%--------------------------------------------------------------------------------------------------%
%                                                                                                  %
\end{abstract}
\section{Introduction}
%--------------------------------------------------------------------------------------------------%
\label{sec: Introduction}
%--------------------------------------------------------------------------------------------------%

Quantum information science can be applied to network security in at least two areas. One is the protection of communication by quantum means: key distribution \cite{Bennett1984, Ekert1991, Ampatzis2021, Cao2022, Sun2022, Durreshahwar2024}, entanglement-based protocols for quantum secret sharing \cite{Ampatzis2022, Ampatzis2023, Andronikos2024b}, for agreement among distributed parties when some of them may be dishonest, such as detectable Byzantine agreement \cite{Andronikos2023a} or news verification \cite{Andronikos2024}, and for anonymous and private information exchange \cite{Andronikos2023, Andronikos2023b, Karananou2024, Andronikos2024a, Andronikos2025, Andronikos2025c}.
%--------------------------------------------------------------------------------------------------%
% REFS-CUT: The other is the detection of attacks, by quantum machine-learning (QML) models trained on network traffic \cite{kaissar2026, corli2025, aparcanatasayco2025, lamichhane2025}, and it is the subject of this paper.
%--------------------------------------------------------------------------------------------------%
The other, which is the subject of this paper, concerns the detection of attacks, by quantum machine-learning (QML) models trained on network traffic \cite{Kaissar2026}.
%--------------------------------------------------------------------------------------------------%
% REFS-CUT: Machine-learning and deep-learning intrusion detection systems (IDS) are widely studied \cite{apruzzese2023, halbouni2022, Azam2023, hozouri2025, chinnasamy2025, ali2025} and are commonly evaluated on public datasets such as NSL-KDD \cite{Tavallaee2009, Azam2023, fu2022}.
%--------------------------------------------------------------------------------------------------%
Machine-learning and deep-learning intrusion detection systems (IDS) are widely studied and commonly evaluated on public datasets such as NSL-KDD \cite{Tavallaee2009, Azam2023, Ali2025}.
%--------------------------------------------------------------------------------------------------%
% REFS-CUT: They recognize attack types present in their training data with few false alarms, whereas some attack types absent from training are not detected \cite{zoppi2023, sarhan2023}; detectors that adapt to new attack types are an active research topic \cite{dimonda2024, li2025}.
%--------------------------------------------------------------------------------------------------%
They recognize attack types present in their training data with few false alarms, whereas some attack types absent from training are not detected \cite{Sarhan2023, Zoppi2023, DiMonda2024}. Among QML detectors, variational quantum classifiers (henceforth VQCs for short), parameterized circuits whose weights are trained by a classical optimizer \cite{Havlicek2019, Cerezo2021}, have been applied together with quantum support-vector machines, quantum neural networks and hybrid detectors, in simulation and on quantum hardware (subsection~\ref{subsec: QML for Intrusion Detection}, Table~\ref{tbl: Related Studies}). A survey of the studies published up to January 2026 finds that most of them use simulated quantum environments and legacy datasets with inconsistent evaluation practices, and calls for standardized evaluation protocols \cite{Kaissar2026}.

Controlled benchmarks provide such protocols for QML classifiers outside intrusion detection.
Bowles et al. tested 12 quantum classifiers on 160 datasets derived from 6 binary classification tasks \cite{Bowles2024}, Schnabel et al.\ trained more than 20,000 quantum-kernel models on 64 datasets \cite{Schnabel2025}, and further studies compare quantum kernels with support-vector machines and logistic regression across feature maps and hyperparameters \cite{Alvarezestevez2025, Egginger2024}. On classical data, these studies find that classical models currently reach the same or higher accuracy \cite{Slattery2023, Bowles2024, Alvarezestevez2025, Florezablan2025}. They also identify the design choices that bring the quantum models closer: with a tuned bandwidth quantum kernels generalize \cite{Canatar2022}, their accuracy improves as qubits are added, and becomes competitive with the best classical methods \cite{Shaydulin2022}. A a careful choice of the feature map and its hyperparameters may be more effective than additional kernel training \cite{Alvarezestevez2025}.
%--------------------------------------------------------------------------------------------------%
% REFS-CUT: Theoretical and numerical work has mapped which variational models admit classical surrogates or efficient classical simulation \cite{schreiber2023, shin2024, Sweke2025, cerezo2025, bermejo2026}, identified properties that a quantum advantage requires \cite{Sweke2025, kubler2021}, and proved, under standard complexity-theoretic assumptions, a learning advantage for quantum models that are deployed classically after training \cite{jerbi2024}.
%--------------------------------------------------------------------------------------------------%
Theoretical and numerical work has mapped which variational models admit classical surrogates \cite{Schreiber2023, Sweke2025} and identified properties that a quantum advantage requires \cite{Sweke2025, Kubler2021}. Reviews treat the conditions for a practical advantage on classical data as an open question \cite{Cerezo2022, Gujju2024, Wang2024} and argue that the focus on outperforming classical models should be complemented by other research perspectives \cite{Schuld2022}. The conditions for a practical advantage on classical data remain an open question, and it has been argued that the focus on outperforming classical models should be complemented by other research perspectives \cite{Schuld2022}.

This study investigates which components limit the accuracy of variational quantum classifiers (VQCs) and quantifies their respective impacts. A VQC consists of four core components \cite{Benedetti2019, Schuld2020, Cerezo2021, LiWeikang2022}:

%--------------------------------------------------------------------------------------------------%
\begin{itemize}
	%----------------------------------------------------------------------------------------------%
	% [ left = 0.00 cm, labelsep = 0.55 cm ]
	%----------------------------------------------------------------------------------------------%
	\item	%	[\textcolor{WordAquaAccent1Darker25} \faListUl]
	The encoding, which maps a data record to a quantum state.
	%----------------------------------------------------------------------------------------------%
	\item	%	[\textcolor{WordAquaAccent1Darker25} \faListUl]
	The ansatz, which is a parameterized unitary.
	%----------------------------------------------------------------------------------------------%
	\item	%	[\textcolor{WordAquaAccent1Darker25} \faListUl]
	The classical optimizer, which trains the parameters of the ansatz.
	%----------------------------------------------------------------------------------------------%
	\item	%	[\textcolor{WordAquaAccent1Darker25} \faListUl]
	The readout, which maps measurement outcomes to a class.
	%----------------------------------------------------------------------------------------------%
\end{itemize}
%--------------------------------------------------------------------------------------------------%

Prior work has evaluated encoding type, circuit depth, and readout as key determinants of classification performance in quantum neural networks \cite{Nguyen2022}. The encoding determines the decision boundaries a classifier can learn \cite{LaRose2020} as well as the frequencies of the functions it can represent \cite{Schuld2021}. Across two datasets, circuits sharing the same topology can vary in accuracy by 10 to 30 percentage points depending solely on the choice of encoding \cite{Tudisco2026}. Furthermore, the encoding choice significantly influences the accuracy of hybrid detectors on network-attack data \cite{Kadi2025}. Because the ansatz is unitary, it preserves the overlaps between encoded states; consequently, any class information discarded during encoding cannot be recovered at any circuit depth \cite{Schuld2021, Thomos2026}. Amplitude encoding stores $2^{ n }$ features in $n$ qubits, at the cost of normalizing the data record and requiring a state-preparation circuit whose size scales with the number of features \cite{Mottonen2005, Shende2006, Plesch2011, Ashhab2022, Nakaji2022, Sun2023}. Under general data assumptions, its encoded states concentrate toward a common state, establishing a lower bound on the training loss \cite{WangXin2025}. In contrast, angle encoding uses one qubit and one rotation per feature while preserving the scale of each feature \cite{LaRose2020, Ranga2024, Rath2024}. An ansatz is characterized by its expressibility \cite{sim2019}, but training can be hindered by challenges such as barren plateaus and local minima \cite{McClean2018, Anschuetz2022, Ragone2024, Cunningham2025}. Most current QML models use a measurement protocol fixed in advance rather than chosen for the task \cite{Chen2025}. Readouts designed for the specific task have been reported to raise accuracy: learnable observables \cite{Chen2025}, measurement schemes for multi-class outputs \cite{Zeng2022, Baek2023} and noise-aware processing of the measurement outcomes on quantum hardware \cite{WangHanrui2022}. The choice of observable also fixes the family of functions that the circuit can represent \cite{Schuld2021b, Jerbi2023}.

\textbf{Contribution.}
This work builds upon the strengths of prior research \cite{Thomos2026}, and offers the following contributions.

%--------------------------------------------------------------------------------------------------%
\begin{itemize}
	%----------------------------------------------------------------------------------------------%
	% [ left = 0.00 cm, labelsep = 0.55 cm ]
	%----------------------------------------------------------------------------------------------%
	\item	[\textcolor{WordAquaAccent1Darker25} \faListUl]
	\textbf{Angle versus amplitude encoding.}
	Neither is better overall; the choice depends on the expected traffic and on the hardware. On traffic similar to the training data, angle encoding gives the more accurate detector, $1.6$ to $4.5$ percentage points above amplitude encoding with the same number of parameters and a trained readout. When attacks of new types are expected, amplitude encoding is the better choice: it detects more than half of the attacks whose type never appeared in training, against a third to less than half for angle encoding, and the classical references split the same way, so the difference comes from the encodings, not from training. On hardware, amplitude encoding saves qubits (4 instead of 16 for 16 features), whereas from 8 features up angle encoding saves two-qubit gates and more than half of the depth.
	%----------------------------------------------------------------------------------------------%
	\item	[\textcolor{WordAquaAccent1Darker25} \faListUl]
	\textbf{Classical versus quantum methods.}
	On the same inputs, the quantum circuits perform at the level of classical models, and the experiments show no quantum advantage. The best circuits are $0.4$ to $1.0$ points more accurate than logistic regression but 0.5 to 1.9 points less accurate than a classical support-vector machine on their own kernel, and every circuit read by a fixed parity is less accurate than logistic regression. Amplitude-encoded circuits do detect more unseen attacks than logistic regression, but so does the classical support-vector machine on the same encoding: the gain comes from the encoding, which a classical computer evaluates in closed form, not from quantum computation. Whether a quantum encoding can do better is a question for encodings whose kernels classical computers cannot evaluate efficiently \cite{Havlicek2019, HuangHY2021}.
	%----------------------------------------------------------------------------------------------%
	\item	[\textcolor{WordAquaAccent1Darker25} \faListUl]
	\textbf{Where the accuracy is lost.}
	in the readout. For both encodings, most of the accuracy is lost at the readout, the step that turns measurement outcomes into a class. A circuit read by the parity of all qubits stays $2.6$ to $7.4$ points below its attainable reference, and the earlier classifier, which reads a single qubit, $11.4$ points; a linear readout trained together with the circuit recovers most of this distance, from the same measurements and with the same tolerance to a finite number of shots.
	%----------------------------------------------------------------------------------------------%
	\item	[\textcolor{WordAquaAccent1Darker25} \faListUl]
	\textbf{Depth versus readout.}
	It is preferable to improve the readout before adding layers. The accuracy plateau over depth reported in \cite{Thomos2026} is a plateau of its parity readout, confirming the readout bottleneck that the study pointed to. With a readout trained together with the circuit, accuracy is higher at every depth, an 8-parameter circuit outperforms the 72-parameter circuit read by parity ($0.917$ against $0.882$), and further depth adds only about two points.
	%----------------------------------------------------------------------------------------------%
	\item	[\textcolor{WordAquaAccent1Darker25} \faListUl]
	\textbf{A yardstick for new classifiers.}
	For each encoding, a classical support-vector machine on the overlaps of the encoded states estimates the accuracy that a well-trained circuit with a linear readout can reach. Reporting a trained circuit as its distance from this reference separates what the encoding allows from what the circuit leaves unused, and for the encodings studied here the reference needs no quantum simulation, so it can be computed before any circuit is trained.
	%----------------------------------------------------------------------------------------------%
\end{itemize}
%--------------------------------------------------------------------------------------------------%

%--------------------------------------------------------------------------------------------------%
%                                                                                                  %
%			W h e r e   D o e s   A                                                                %
%			V a r i a t i o n a l   Q u a n t u m   C l a s s i f i e r                            %
%			L o s e   A c c u r a c y   ?                                                          %
%                                                                                                  %
%--------------------------------------------------------------------------------------------------%
%                                                                                                  %
%			S u b s e c t i o n                                                                    %
%                                                                                                  %
%--------------------------------------------------------------------------------------------------%
%                                                                                                  %
%			O r g a n i z a t i o n                                                                %
%                                                                                                  %
%--------------------------------------------------------------------------------------------------%
\subsection*{Organization}
%--------------------------------------------------------------------------------------------------%
\label{subsec: Organization}
%--------------------------------------------------------------------------------------------------%

This article is organized in the following way. Section \ref{sec: Introduction} introduces the topic and provides references to relevant literature. Section~\ref{sec: Background} provides a comprehensive and succinct reminder of the basic concepts. Section~\ref{sec: Methods and Technologies} describes the methods used in enough detail to trace them to the published code. Section~\ref{sec: Results} reports the results, while Sections~\ref{sec: Discussion} and~\ref{sec: Conclusions} analyze their interpretation and offer an insightful discussion.

%--------------------------------------------------------------------------------------------------%
%                                                                                                  %
%			W h e r e   D o e s   A                                                                %
%			V a r i a t i o n a l   Q u a n t u m   C l a s s i f i e r                            %
%			L o s e   A c c u r a c y   ?                                                          %
%                                                                                                  %
%--------------------------------------------------------------------------------------------------%
%                                                                                                  %
%			S e c t i o n   2                                                                      %
%                                                                                                  %
%--------------------------------------------------------------------------------------------------%
%                                                                                                  %
%			B a c k g r o u n d                                                                    %
%                                                                                                  %
%--------------------------------------------------------------------------------------------------%
\section{Background}
%--------------------------------------------------------------------------------------------------%
\label{sec: Background}
%--------------------------------------------------------------------------------------------------%

%--------------------------------------------------------------------------------------------------%
%                                                                                                  %
%			W h e r e   D o e s   A                                                                %
%			V a r i a t i o n a l   Q u a n t u m   C l a s s i f i e r                            %
%			L o s e   A c c u r a c y   ?                                                          %
%                                                                                                  %
%--------------------------------------------------------------------------------------------------%
%                                                                                                  %
%			S u b s e c t i o n   2 . 1                                                            %
%                                                                                                  %
%--------------------------------------------------------------------------------------------------%
%                                                                                                  %
%			Q M L   F o r   I n t r u s i o n   D e t e c t i o n                                  %
%                                                                                                  %
%--------------------------------------------------------------------------------------------------%
\subsection{Quantum Machine Learning for Intrusion Detection}
%--------------------------------------------------------------------------------------------------%
\label{subsec: QML for Intrusion Detection}
%--------------------------------------------------------------------------------------------------%

Table~\ref{tbl: Related Studies} summarizes the quantum intrusion-detection studies cited here, as each reports itself; their datasets, models and metrics differ. One study executed a reduced version of its architecture on a trapped-ion computer \cite{Kukliansky2024}, and most of this literature is evaluated in simulation \cite{Kaissar2026}. The studies that isolate one factor point in the direction taken here: the encoding changes the accuracy of otherwise identical hybrid detectors \cite{Kadi2025}, quantum kernels on IoT data perform comparably to classical ones \cite{Damore2026}, and in malicious-URL detection a quantum SVM outperformed the classical SVM while a classical convolutional network outperformed its quantum counterpart \cite{Eze2025}.
%--------------------------------------------------------------------------------------------------%
% REFS-CUT: Surveys map these models \cite{corli2025}, describe them as promising \cite{aparcanatasayco2025, lamichhane2025} and call for standardised evaluation protocols \cite{Kaissar2026}.
%--------------------------------------------------------------------------------------------------%
In the control study \cite{Thomos2026}, accuracy saturated at 0.880 over five depths while the training objective kept falling, so depth was not the binding constraint; the encoding and the readout, held fixed there, are varied separately here.

%--------------------------------------------------------------------------------------------------%
\begin{table}[!htb]
	\scriptsize
	%----------------------------------------------------------------------------------------------%
	\caption{Quantum intrusion-detection studies cited in this paper, as reported in each study's own summary. ``n.s.'': not stated in the summary.
	Accuracy values are those reported by the authors and are not comparable across rows, because datasets, splits and metrics differ.}
	\label{tbl: Related Studies}
	%----------------------------------------------------------------------------------------------%
	\centering
	%----------------------------------------------------------------------------------------------%
	\SetTblrInner { rowsep = 0.300 mm }
	%----------------------------------------------------------------------------------------------%
	\begin{tblr}
		{
			colspec =
			{
				Q [ l, m, 1.500 cm ]
				| [ 0.500 pt, WordAquaDarker25 ]
				X [ 1.10, l, m ]
				| [ 0.500 pt, WordAquaDarker25 ]
				X [ 1.20, l, m ]
				| [ 0.500 pt, WordAquaDarker25 ]
				Q [ c, m, 1.500 cm ]
				| [ 0.500 pt, WordAquaDarker25 ]
				X [ 1.40, l, m ]
			},
			row{1} = { bg = WordAquaDarker25, fg = white, font = \bfseries },
			hline{1, Z} = { 1.000 pt, WordAquaDarker25 },
			hline{2} = { 0.750 pt, WordAquaDarker25 },
			hline{3-Y} = { 0.250 pt, WordAquaLighter60 },
			row{even[2]} = { bg = WordAquaLighter80 },
		}
		%------------------------------------------------------------------------------------------%
		Study & Data & Quantum model & Execution & Reported outcome \\
		%------------------------------------------------------------------------------------------%
		% REFS-CUT: \cite{kalinin2022} & Own packet-stream dataset & QSVM, QCNN & n.s. & 98\% accuracy, twice as fast as classical detectors \\
		% REFS-CUT: \cite{islam2022} & In-vehicle CAN bus & Hybrid quantum--classical network & n.s. & 94\%; LSTM 88\%, quantum network alone 62\% \\
		% REFS-CUT: \cite{salek2023} & In-vehicle CAN bus (fuzzy attacks) & Classical network + quantum RBM & n.s. & 97.0--98.3\%; classical-only 86.7--95\% \\
		% REFS-CUT: \cite{kumar2025} & IoT traffic & QSVM & n.s. & n.s. \\
		% REFS-CUT: \cite{elsedimy2024} & IoT traffic & QSVM tuned by a grey-wolf optimiser & n.s. & Better than recent detection models \\
		% REFS-CUT: \cite{qi2024} & KDD Cup 99, UNSW-NB15 & Variational quantum CNN & n.s. & Smaller accuracy loss under FGSM attacks \\
		% REFS-CUT: \cite{kucukkara2024} & CIC-DDoS2019 (subset) & Quantum neural network & Simulation & 92.63\% \\
		%------------------------------------------------------------------------------------------%
		\cite{Kim2024} & Own DDoS traffic & Angle embedding, fidelity and entropy outlier analysis & n.s. & 99.87\% \\
		%------------------------------------------------------------------------------------------%
		% REFS-CUT: \cite{hdaib2024} & Computer and IoT flows & Quantum autoencoder + one-class SVM, random forest or kNN & n.s. & Best with quantum kNN \\
		% REFS-CUT: \cite{abreu2024} & Public datasets & Hybrid QML detector & n.s. & Outperforms classical ML \\
		%------------------------------------------------------------------------------------------%
		\cite{Kukliansky2024} & NF-UNSW-NB15 & Multi-layer quantum neural network & IonQ Aria-1 & F1 = 0.86 \\
		%------------------------------------------------------------------------------------------%
		% REFS-CUT: \cite{rosaremedios2024} & Several cybersecurity datasets & Known QML models, parameterisations compared & n.s. & Promising against classical models \\
		%------------------------------------------------------------------------------------------%
		\cite{Eze2025} & Malicious URLs & QSVM, QCNN (ZZ- and Z-feature maps) & n.s. & QSVM $>$ SVM; CNN $>$ QCNN \\
		\cite{Kadi2025} & Two network-attack datasets & Amplitude, angle, IQP and QAOA embeddings & n.s. & Accuracy depends on the encoding \\
		% REFS-CUT: \cite{sarvade2025} & UNB-CIC-DDoS, Kaggle DDoS & Hybrid quantum neural network (PennyLane) & n.s. & 99\% and 90\% \\
		\cite{Damore2026} & IoT room occupancy & Projected and fidelity quantum kernels & n.s. & Comparable to classical methods \\
		\cite{Thomos2026} & NSL-KDD (20\% subset) & Amplitude-encoded VQC, 4 qubits & Simulation & 88\%; saturates over depth \\
		%------------------------------------------------------------------------------------------%
		\textbf{This work} & NSL-KDD (20\% subset) & Amplitude- and angle-encoded VQCs, 2--16 qubits, three readouts & Exact simulation & Section~\ref{sec: Results} \\
		%------------------------------------------------------------------------------------------%
		\end{tblr}
	%----------------------------------------------------------------------------------------------%
\end{table}
\subsection{Encodings, Kernels \& Provable Quantum Classification}
%--------------------------------------------------------------------------------------------------%
\label{subsec: Encodings and Kernels}
%--------------------------------------------------------------------------------------------------%

\noindent\textbf{Encodings.}
With a single layer of rotations a quantum model is a truncated Fourier series whose frequencies are set by the encoding \cite{Schuld2021}, and re-uploading the data enlarges that set \cite{PerezSalinas2020}.
%--------------------------------------------------------------------------------------------------%% REFS-CUT: Encodings differ in robustness to noise \cite{LaRose2020}, can be searched for automatically \cite{Nguyen2022b}, and together with their pre-processing and hyperparameters change the accuracy of quantum classifiers substantially \cite{john2023, egginger2024}.
%--------------------------------------------------------------------------------------------------%
Encodings differ in robustness to noise \cite{LaRose2020} and, together with their pre-processing and hyperparameters, may affect the accuracy of quantum classifiers substantially \cite{Egginger2024}.
%--------------------------------------------------------------------------------------------------%
% REFS-CUT: Exact amplitude encoding of $n$ qubits needs a number of two-qubit gates that grows exponentially with $n$ \cite{Mottonen2005, Shende2006, Plesch2011, iten2016}, a cost that can be traded against ancillas and depth \cite{Sun2023} or lowered by approximate and probabilistic schemes \cite{Ashhab2022, Nakaji2022}.
%--------------------------------------------------------------------------------------------------%
Exact amplitude encoding of $n$ qubits requires a number of two-qubit gates that grows exponentially with $n$ \cite{Mottonen2005, Shende2006, Plesch2011}, a cost that can be traded against ancillas and depth \cite{Sun2023}.

\noindent\textbf{Kernels \& surrogates.}
%--------------------------------------------------------------------------------------------------%
% REFS-CUT: A model that encodes once and measures an observable is a kernel method whose kernel is the fidelity between encoded states \cite{Schuld2021b, schuld2019, Havlicek2019}; such kernels can be trained on hardware \cite{hubregtsen2022} and their expressivity is well characterised \cite{GilFuster2024b}.
%--------------------------------------------------------------------------------------------------%
A model that encodes once and measures an observable is a kernel method whose kernel is the fidelity between encoded states \cite{Havlicek2019, Schuld2021b}. On classical data their values concentrate as qubits are added, and they generalize only once their bandwidth is reduced \cite{Kubler2021, Canatar2022, Shaydulin2022, Thanasilp2024}, at which point they resemble classical kernels \cite{Slattery2023, Florezablan2025}. Whether a quantum kernel can help depends on the data and can be checked in advance \cite{HuangHY2021}.
%--------------------------------------------------------------------------------------------------%
% REFS-CUT: Explicit models can go beyond kernels \cite{Jerbi2023}; their generalisation is bounded by their trainable gates \cite{caro2022}, although such bounds do not explain it in practice \cite{GilFuster2024}.
%--------------------------------------------------------------------------------------------------%
Explicit models can go beyond kernels \cite{Jerbi2023}.
%, and bounds on their generalisation by the number of trainable gates do not explain it in practice \cite{GilFuster2024}.
In this work, the kernel SVM is used only as a measured reference for one encoding.

\noindent\textbf{Provable classification \& hardware.}
Variational classifiers are heuristic. On real devices, noise flattens training landscapes \cite{Wang2021}, error mitigation reduces bias at a sampling cost \cite{Cai2023} without restoring trainability in general. Readout errors can be partly corrected and benchmarks of annealers and hybrid solvers show that the formulation of a problem for the hardware sets its qubit count and the validity of its solutions \cite{Papalitsas2019, Stogiannos2022}. A complementary line of work designs quantum classifiers with proven behavior: a single-query oracle algorithm classifies a hierarchy of classes of imbalanced Boolean functions with probability $1.0$ \cite{Andronikos2025a}, still returns useful information for functions close to the promised class \cite{Andronikos2025b}, and has a success probability that falls, with systematic exceptions, within stated intervals of the Hamming distance from that class \cite{Andronikos2026}. This paper asks the same question, what a classifier can recover from its input, empirically for a variational classifier on network traffic.
\section{Methods \& Technologies}
%--------------------------------------------------------------------------------------------------%
\label{sec: Methods and Technologies}
%--------------------------------------------------------------------------------------------------%

\begin{figure}[!htb]
	%----------------------------------------------------------------------------------------------%
	\centering
	%----------------------------------------------------------------------------------------------%
	\begin{tikzpicture}
		[
			blk/.style =
			{
				rectangle, rounded corners = 3pt, draw = #1!80!black, fill = #1!10,
				line width = 0.8pt, align = center, font = \small\sffamily,
				minimum height = 1.00cm, text width = 5.80cm, inner sep = 3pt
			},
			side/.style = { blk = #1, text width = 4.40cm, minimum height = 1.40cm },
			db/.style =
			{
				cylinder, shape border rotate = 90, aspect = 0.15, draw = #1!80!black, fill = #1!10,
				line width = 0.8pt, align = center, font = \small\sffamily,
				minimum width = 6.00cm, minimum height = 1.40cm, inner sep = 3pt
			},
			arr/.style = { -{Stealth[length = 2.4mm]}, line width = 0.9pt, WordBlueDarker50 },
			lbl/.style = { font = \footnotesize\sffamily, text = WordBlueDarker50 }
		]
		%------------------------------------------------------------------------------------------%
		%                                                                                          %
		%			Shared steps (centre), the two measured paths                                  %
		%			(left: trained circuit; right: reference)                                      %
		%                                                                                          %
		%------------------------------------------------------------------------------------------%
		\node [ db = WordBlueDark ] (data) at (0.00, 0.00) { \textbf{NSL-KDD}\\20\% subset, KDDTest+ };
		\node [ blk = WordBlueDark ] (prep) at (0.00, -1.85) { \textbf{Preprocessing}\\target encoding, scaling, PCA };
		\node [ blk = MagentaDark ] (enc) at (0.00, -3.55) { \textbf{Encoding}\\amplitude or angle };
		\node [ side = WordAquaDarker25 ] (ans) at (-2.85, -6.30) { \textbf{Ansatz}\\$U(\boldsymbol{\theta})$: $R_{y}$ + CX };
		\node [ side = WordAquaDarker25 ] (ro) at (-2.85, -8.30) { \textbf{Readout}\\qubit-0 or global parity\\or linear head };
		\node [ side = WordGoldAccent1Darker25 ] (ker) at (2.85, -6.30) { \textbf{Fidelity kernel}\\$F(\mathbf{x}, \mathbf{y})$ };
		\node [ side = WordGoldAccent1Darker25 ] (svm) at (2.85, -8.30) { \textbf{SVM}\\precomputed $F$\\$C \in \{1, 100\}$ };
		\node [ blk = RedPurple, text width = 6.80cm ] (cmp) at (0.00, -10.50) { \textbf{Compare}\\trained circuit vs.\ attainable reference };
		%------------------------------------------------------------------------------------------%
		\draw [arr] (data) -- (prep);
		\draw [arr] (prep) -- node [ lbl, right = 2pt ] {$\mathbf{x} \in \mathbb{R}^{N}$} (enc);
		\draw [ line width = 0.9pt, WordBlueDarker50 ] (enc.south) -- node [ lbl, right = 2pt ] {$\ket{\phi(\mathbf{x})}$} ++(0, -0.70) coordinate (split);
		\draw [arr] (split) -| node [ lbl, pos = 0.75, left = 3pt, align = right ] {trained\\by COBYLA} (ans.north);
		\draw [arr] (split) -| node [ lbl, pos = 0.75, right = 3pt, align = left ] {closed form,\\no simulation} (ker.north);
		\draw [arr] (ans) -- node [ lbl, right = 2pt ] {$p(b)$} (ro);
		\draw [arr] (ker) -- (svm);
		\draw [arr] (ro.south) -- ++(0, -0.45) -| ($ (cmp.north west)!0.25!(cmp.north east) $);
		\draw [arr] (svm.south) -- ++(0, -0.45) -| ($ (cmp.north west)!0.75!(cmp.north east) $);
	\end{tikzpicture}
	%----------------------------------------------------------------------------------------------%
	\caption{The two paths measured for every configuration.
	Left path: a variational circuit trained with one of three readouts.
	Right path: a support-vector machine on the fidelity kernel of the same encoded states, whose accuracy is the attainable reference of the encoding (Section~\ref{subsec: Reference}).
	Both paths use the same training records and are scored on the same test records.}
	\label{fig: Pipeline}
	%----------------------------------------------------------------------------------------------%
\end{figure}
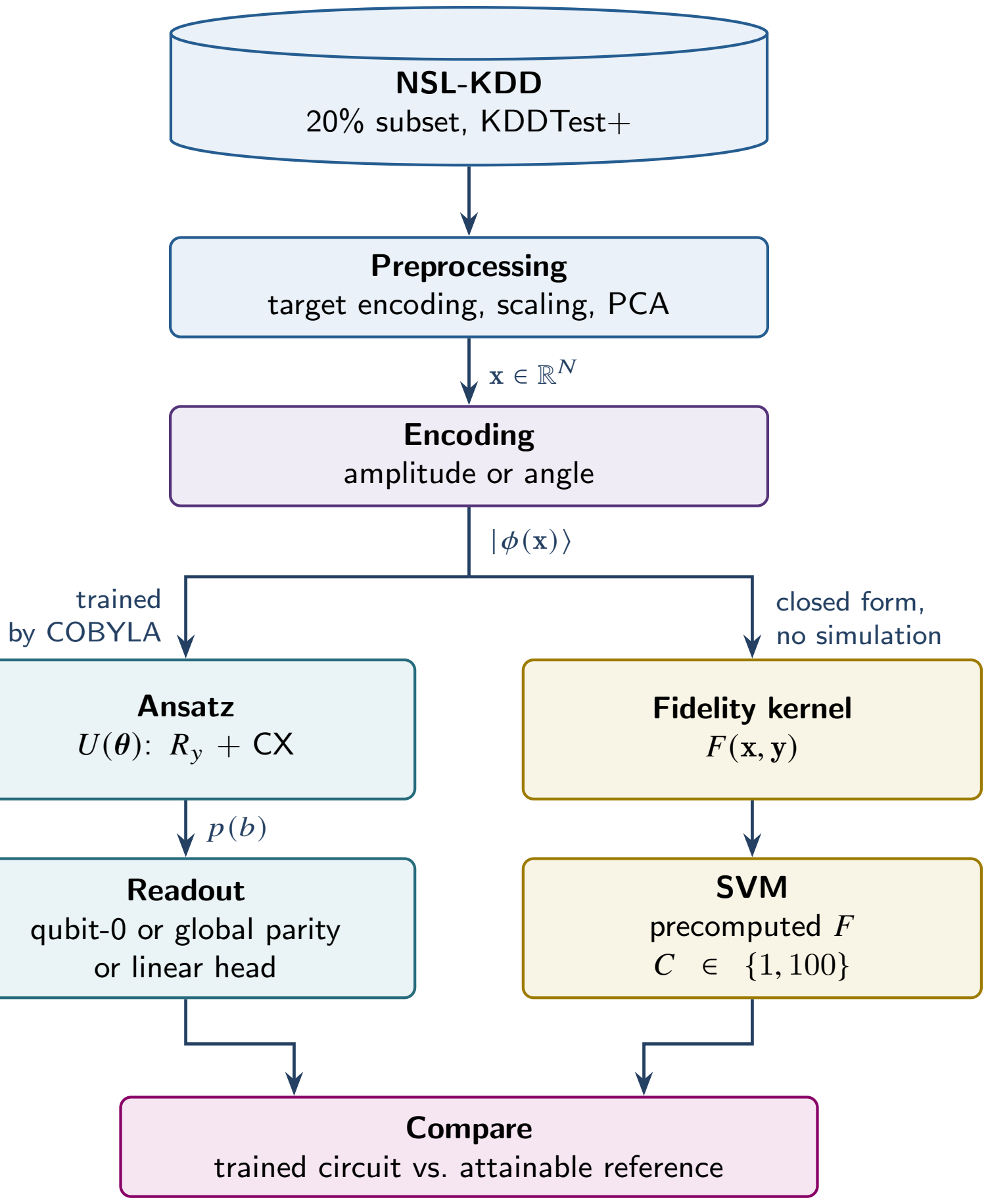

Figure~\ref{fig: Pipeline} shows the pipeline. Every record passes through the same preprocessing, is encoded into a quantum state, and follows two paths: a trained circuit with a readout, and the fidelity-kernel SVM that gives the attainable reference of its encoding. Comparing the two paths on the same records is the measurement this paper makes. Sections~\ref{subsec: Data}--\ref{subsec: Circuits} describe each block and what is inherited from the control study, Section~\ref{subsec: Simulation} the techniques that make full-scale runs affordable and Section~\ref{subsec: Design} the experiments.

%--------------------------------------------------------------------------------------------------%
\subsection{Data, Preprocessing \& Encodings}
%--------------------------------------------------------------------------------------------------%
\label{subsec: Data}
%	\label{subsec: Encodings}
%--------------------------------------------------------------------------------------------------%

As in the control study \cite{Thomos2026}, the official 20\% subset of NSL-KDD \cite{Tavallaee2009} is split 80/20, stratified on the label crossed with a band of the payload size (20,153 training and 5,039 internal test records).
%--------------------------------------------------------------------------------------------------%
% REFS-CUT: The three categorical fields (protocol, service, flag) and the payload band are target-encoded \cite{Miccibarreca2001} on the binary label of the training records (normal = 0, attack = 1), all features are standardised, and 16 principal components (PCs) are retained \cite{jolliffe2002}, every transformation being fitted on the training partition; the 16 PCs retain 0.8550 of the variance (0.8678 in \cite{Thomos2026}).
%--------------------------------------------------------------------------------------------------%
The three categorical fields (protocol, service, flag) and the payload band are target-encoded \cite{Miccibarreca2001} on the binary label of the training records (normal = 0, attack = 1), all features are standardized, and 16 principal components (PCs) are retained, every transformation being fitted on the training partition; the 16 PCs retain 0.8550 of the variance (0.8678 in \cite{Thomos2026}). Evaluation uses the internal test partition, the complete KDDTest+ file (22,544 records), and the 3,752 KDDTest+ attack records whose type is absent from the training file (``novel''), on which recall is reported. The control study is re-run with this preprocessing, so its numbers here are comparable with the published ones but not identical; Section~\ref{subsec: Anchor} compares the two.

With $N$ principal components $\mathbf{x} = (x_{1}, \dots, x_{N})$, $\ell_{2}$ amplitude encoding normalizes the vector and loads it into the amplitudes of $\lg N$ qubits, $\ket{ \phi_{ \mathrm{amp} } ( \mathbf{x} ) } = \sum_{i} \hat{x}_{i} \ket{i}$ with $\hat{\mathbf{x}} = \mathbf{x} / \norm{ \mathbf{ x } }$. Angle encoding uses one qubit per component and rotates qubit $i$ by $R_{y} ( s x_{i} )$, where $s$ is a constant that scales the components into angles. A second variant (angle-phase) applies to each qubit a Hadamard gate followed by a phase rotation of $2 s x_{i}$.
% one repetition of the \texttt{ZFeatureMap} of Qiskit \cite{JavadiAbhari2024}, the single-qubit ($Z$-only) case of the feature map of \cite{Havlicek2019}.
The control study \cite{Thomos2026} used this map in its separability probe and in its four-qubit trained probe, while its classifier used amplitude encoding. The fidelity of two encoded states, $F ( \mathbf{ x }, \mathbf{ y } ) = \abs{\braket{\phi(\mathbf{x})}{\phi(\mathbf{y})}}^{2}$, has a closed form for all three encodings:

%--------------------------------------------------------------------------------------------------%
\begin{align}
	\hspace{ - 1.000 cm }
	F_{\mathrm{amp}}(\mathbf{x}, \mathbf{y}) = (\hat{\mathbf{x}} \cdot \hat{\mathbf{y}})^{2},
	\quad
	F_{\mathrm{RY}}(\mathbf{x}, \mathbf{y}) = \prod_{i = 1}^{N} \cos^{2} \frac{ s (x_{i} - y_{i}) }{ 2 },
	\quad
	F_{\mathrm{phase}}(\mathbf{x}, \mathbf{y}) = \prod_{i = 1}^{N} \cos^{2} \big( s (x_{i} - y_{i}) \big).
	\label{eq: Fidelity Kernels}
\end{align}
%--------------------------------------------------------------------------------------------------%

Equation~\eqref{eq: Fidelity Kernels} shows what each encoding keeps. Amplitude encoding keeps only the direction of the record and compares directions through a squared cosine, a homogeneous quadratic kernel; the norm of the record is lost. Angle encoding keeps every component and its scale, and for small $s$ its kernel approaches a Gaussian kernel with width set by $s$ \cite{Shaydulin2022, Florezablan2025}. The scale $s = 0.250$ was chosen from $\{ 0.25, 0.5, 1, 2, \pi \}$ as the value at which the pairwise separability of angle-phase states at $N = 16$ comes closest to the value published in \cite{Thomos2026} (AUC 0.762); agreement with that value is therefore a consequence of the choice and not an independent check.

\noindent\textbf{Separability.} The separability AUC uses the fidelity of a pair of records as a score that the two records share a class

%--------------------------------------------------------------------------------------------------%
\begin{align}
	\mathrm{AUC} = P \big( F_{\mathrm{same}} > F_{\mathrm{diff}} \big) + \tfrac{1}{2} P \big( F_{\mathrm{same}} = F_{\mathrm{diff}} \big),
	\label{eq: AUC}
\end{align}
%--------------------------------------------------------------------------------------------------%

estimated on 200,000 random pairs of training records. Because every ansatz is unitary and preserves inner products, this quantity is the same before and after the circuit, at any depth.
%; it is, however, not on the scale of accuracy, which is what the attainable reference provides.

%--------------------------------------------------------------------------------------------------%
%                                                                                                  %
%			W h e r e   D o e s   A                                                                %
%			V a r i a t i o n a l   Q u a n t u m   C l a s s i f i e r                            %
%			L o s e   A c c u r a c y   ?                                                          %
%                                                                                                  %
%--------------------------------------------------------------------------------------------------%
%                                                                                                  %
%			S u b s e c t i o n   3 . 2                                                            %
%                                                                                                  %
%--------------------------------------------------------------------------------------------------%
%                                                                                                  %
%			T h e   A t t a i n a b l e   R e f e r e n c e                                        %
%                                                                                                  %
%--------------------------------------------------------------------------------------------------%
\subsection{The Attainable Reference}
%--------------------------------------------------------------------------------------------------%
\label{subsec: Reference}
%--------------------------------------------------------------------------------------------------%

A circuit that encodes the record once, applies a unitary $U ( \boldsymbol{ \theta } )$ and measures an observable $O$, in effect computes
%--------------------------------------------------------------------------------------------------%
\begin{align}
	f_{\boldsymbol{\theta}}(\mathbf{x})
	=
	\bra{\phi(\mathbf{x})} U^{\dagger}(\boldsymbol{\theta}) \, O \, U(\boldsymbol{\theta}) \ket{\phi(\mathbf{x})}
	=
	\operatorname{Tr}\big[ M_{\boldsymbol{\theta}} \, \rho(\mathbf{x}) \big]
	\ ,
	\quad \rho(\mathbf{x}) = \ket{\phi(\mathbf{x})} \bra{\phi(\mathbf{x})}
	\ ,
	\label{eq: Linear Model}
\end{align}
%--------------------------------------------------------------------------------------------------%

where $M_{\boldsymbol{\theta}} = U^{\dagger}(\boldsymbol{\theta}) \, O \, U(\boldsymbol{\theta})$ is linear in
% the encoded state
$\rho(\mathbf{x})$, for every choice of $\boldsymbol{\theta}$ and $O$ \cite{Schuld2021b}.

A linear readout of several observables, $\sum_{S} w_{S} \langle Z_{S} \rangle + b$, has the same form with $O = \sum_{S} w_{S} Z_{S}$ and assigns the class by the sign of $f_{\boldsymbol{\theta}}(\mathbf{x}) + b$. The inner product between two encoded states in this space is $\operatorname{Tr}[\rho(\mathbf{x}) \rho(\mathbf{y})] = F(\mathbf{x}, \mathbf{y})$, so a support-vector machine with the precomputed kernel $F$ searches the whole family of such linear models with a convex optimiser: its decision function $\sum_{j} \alpha_{j} y_{j} F(\mathbf{x}_{j}, \mathbf{x}) + b$ is $\operatorname{Tr}[M \rho(\mathbf{x})] + b$ with $M = \sum_{j} \alpha_{j} y_{j} \, \rho(\mathbf{x}_{j})$. Its test accuracy estimates what a circuit with this encoding and a linear readout could reach with an ideal ansatz, an ideal optimiser and an ideal readout: the \emph{attainable reference}.
%--------------------------------------------------------------------------------------------------%
% TONE-CHECK: The difference between the reference and a trained circuit is the accuracy lost to the ansatz, the optimiser and the readout, and not to the encoding.
%--------------------------------------------------------------------------------------------------%
The difference between the reference and a trained circuit therefore estimates the accuracy lost to the ansatz, the optimiser and the readout, rather than to the encoding. The SVM is trained on the training records of the circuit it is compared with and scored on the same test records, with soft margin $C \in \{1, 100\}$. This compares the encodings without circuits and scores it on 2,000 records of the internal partition. The reference is not a strict upper bound, a non-linear function of the outcomes, or a circuit that loads the data more than once; is not linear in $\rho(\mathbf{x})$ and can exceed it \cite{PerezSalinas2020, Jerbi2023}, and a fixed $C$ can leave a fraction of a point unused.

\subsection{Circuits, Readouts, Training \& Protocol}
%--------------------------------------------------------------------------------------------------%
\label{subsec: Circuits}
%	\label{subsec: Provenance}
%--------------------------------------------------------------------------------------------------%

\noindent\textbf{Ansatz.}
The control study used the RealAmplitudes circuit of Qiskit \cite{JavadiAbhari2024}, a hardware-efficient ansatz \cite{Kandala2017}, on four qubits with five repetitions (24 parameters): layers of $R_{y}$ rotations separated by controlled-NOT (CX) gates in a reverse-linear chain, from CX$(n - 2, n - 1)$ to CX$(0, 1)$. The matched comparison of the two encodings (E4, Section~\ref{subsec: Design}) uses one family in both arms, $R_{y}$ layers separated by a linear CX chain, from CX$(0, 1)$ to CX$(n - 2, n - 1)$ (\texttt{n\_local} in Qiskit), with the number of repetitions chosen so that both arms carry $3N$ parameters: two repetitions on $N$ qubits for angle encoding (12, 24 and 48 parameters at $N = 4$, 8 and 16), and 5, 7 and 11 repetitions on 2, 3 and 4 qubits for amplitude encoding.

\noindent\textbf{Readouts.}
A readout maps the measured outcome $b \in \{0, \dots, 2^{n} - 1\}$, or its distribution $p(b)$, to a class. \emph{Qubit-0 parity}, i.e., $c(b) = b \bmod 2$, reads the least significant qubit. This is the default of the \texttt{VQC} class of Qiskit Machine Learning \cite{Sahin2025}, which the control study in \cite{Thomos2026} used. \emph{Global parity}, i.e., $c(b) = \mathrm{popcount}(b) \bmod 2$, reads all qubits; it is the parity function of \cite{Havlicek2019}, $f = Z_{1} Z_{2}$ on its two qubits. A \emph{linear head} is a logistic regression on features of $p(b)$: the 16 outcome probabilities on the four-qubit circuit of the control study, and in E4 the expectations $\langle Z_{S} \rangle = \sum_{b} p(b) (-1)^{\mathrm{popcount}(S \wedge b)}$ of all Pauli-$Z$ strings $S$ of weight one or two plus the global parity, since $Z_{S}$ is diagonal with eigenvalue $(-1)^{\mathrm{popcount}(S \wedge b)}$ on $\ket{b}$. On $n$ qubits these are $n + n (n - 1) / 2 + 1$ features, except on two qubits, where the pair is the global parity: 3, 7 and 11 features for amplitude and 11, 37 and 137 for angle encoding at $N = 4$, 8 and 16. A head can be \emph{frozen}, fitted once on the features of a circuit already trained for a parity readout, or \emph{co-trained}, refitted by logistic regression inside every evaluation of the objective, so that the optimizer changes the circuit while the head follows it. The co-trained head is thus a diagonal observable, $\sum_{b} w_{b} \ket{b} \bra{b}$ or $\sum_{S} w_{S} Z_{S}$, trained together with the circuit parameters, a restricted form of the learnable observables of \cite{Chen2025}. The frozen head measures what a parity readout discards from the states that a given circuit produces; the co-trained head measures what a readout chosen together with the circuit can reach. On the four-qubit amplitude circuit two further readouts complete the range: the best fixed assignment of the 16 outcomes to the two classes, found exhaustively among all $2^{16}$ assignments on the training data, of which qubit-0 parity is one; and a multilayer perceptron (MLP) with 32 hidden units on the outcome probabilities, which is not linear and therefore not covered by the reference.

\noindent\textbf{Training.}
The objective is the cross-entropy of the predicted class probabilities on the training records, minimized with constrained optimization by linear approximation (COBYLA) \cite{powell1994}. Initial parameters are drawn uniformly in $(-0.1, 0.1)$ from the seed, so that within a seed the two arms of E4 start from the same draw and see the same training records. The budget is 400 objective evaluations in E4 and $\max(150, 24(\textit{reps} + 1))$ in the depth sweep, the schedule of \cite{Thomos2026}.

\noindent\textbf{Light cone.}
With a linear CX chain and two repetitions, the expectation of $Z$ on qubit 0 of the angle circuit depends only on the first three components, whose qubits are the only ones inside its backward light cone. A numerical check at $N = 8$ confirms it: changing $x_{0}, \dots, x_{2}$ moves $\langle Z_{0} \rangle$ by 0.443, changing $x_{3}, \dots, x_{7}$ by $3.3 \times 10^{-16}$.
% The qubit-0 readout of the angle arm therefore sees 3 of the $N$ components, and its results are reported (Table~S11) but not used to compare the encodings.

\noindent\textbf{Protocol.}
Table~\ref{tbl: Protocol} compares the protocol of the control study with the present one, row by row, and marks the source of each choice. \emph{From \cite{Thomos2026}} means that the choice in \cite{Thomos2026} is retained, \emph{Re-run} means that a step is computed again with the new code developed for the present study, and \emph{New} that the choice belongs to the present study. Only the rows marked as the independent variable may explain a difference between encodings; the others are measured against the published numbers in Section~\ref{subsec: Anchor}.

%--------------------------------------------------------------------------------------------------%
\begin{table}[!htb]
	%----------------------------------------------------------------------------------------------%
	\footnotesize
	%----------------------------------------------------------------------------------------------%
	\caption{Protocol of the control study \cite{Thomos2026} and of the present work, with the source of each choice.}
	\label{tbl: Protocol}
	%----------------------------------------------------------------------------------------------%
	\centering
	%----------------------------------------------------------------------------------------------%
	\SetTblrInner { rowsep = 0.450 mm }
	%----------------------------------------------------------------------------------------------%
	\begin{tblr}
		{
			colspec =
			{
				Q [ l, m, 2.100 cm ]
				| [ 0.500 pt, WordAquaDarker25 ]
				X [ 1.00, l, m ]
				| [ 0.500 pt, WordAquaDarker25 ]
				X [ 1.45, l, m ]
				| [ 0.500 pt, WordAquaDarker25 ]
				Q [ c, m, 1.500 cm ]
			},
			row{1} = { bg = WordAquaDarker25, fg = white, font = \bfseries },
			hline{1, Z} = { 1.000 pt, WordAquaDarker25 },
			hline{2} = { 0.750 pt, WordAquaDarker25 },
			hline{3-Y} = { 0.250 pt, WordAquaLighter60 },
			cell{2-Z}{4} = { font = \itshape },
		}
		Parameter & Control study & This work & Source \\
		Dataset, split & KDDTrain+\_20Percent, 80/20, stratified on label $\times$ payload band & Same; build of 20,153/5,039 records (Table A2 of \cite{Thomos2026}) & From \cite{Thomos2026} \\
		Preprocessing & Target encoding, standardization, 16 PCs (0.8678 of the variance) & Same steps; target encoder fitted on the binary label of the training records; 16 PCs (0.8550 of the variance) & Re-run \\
		Novel-signature set & 3,752 KDDTest+ records & Same definition & From \cite{Thomos2026} \\
		Simulator & Qiskit state-vector & Exact linear algebra (Equation~\eqref{eq: Amplitude Pass}, Algorithm~\ref{alg: Angle}), verified against Qiskit & New \\
		Training records & 2,000 (stratified subsample) & 2,000 as in \cite{Thomos2026} (E2, E4, E5); full 20,153 (E1, E3) & From \cite{Thomos2026} / New \\
		Encoding & Amplitude, 16 PCs, 4 qubits & Amplitude ($\log_{2} N$ qubits), angle-RY and angle-phase ($N$ qubits), $N = 4, 8, 16$ & Independent variable \\
		Ansatz & RealAmplitudes, reverse-linear CX, 5 reps & Same (E1, E3, E5); $R_{y}$ + linear CX, $3N$ parameters in both arms (E4) & From \cite{Thomos2026} / New \\
		Readout & Qubit-0 parity & Qubit-0 parity, global parity, linear head (frozen and co-trained); best fixed and MLP (E1) & Second factor \\
		Optimizer, budget & COBYLA, $\max(150, 24(\textit{reps} + 1))$ & Same (E1--E3, E5); 400 evaluations (E4) & From \cite{Thomos2026} / New \\
		Reference & Classical baselines & Fidelity-kernel SVM on the same records; logistic regression on $\hat{\mathbf{x}}$ and $\mathbf{x}$ & New \\
		Seeds & 5 & 5; 1 for the 16-qubit angle circuit & From \cite{Thomos2026} \\
		Statistics & Welch and paired $t$, record bootstrap & Same, plus effect sizes and seed intervals & From \cite{Thomos2026} / New \\
	\end{tblr}
	%----------------------------------------------------------------------------------------------%
\end{table}
%--------------------------------------------------------------------------------------------------%

%--------------------------------------------------------------------------------------------------%
%                                                                                                  %
%			W h e r e   D o e s   A                                                                %
%			V a r i a t i o n a l   Q u a n t u m   C l a s s i f i e r                            %
%			L o s e   A c c u r a c y   ?                                                          %
%                                                                                                  %
%--------------------------------------------------------------------------------------------------%
%                                                                                                  %
%			S u b s e c t i o n   3 . 4                                                            %
%                                                                                                  %
%--------------------------------------------------------------------------------------------------%
%                                                                                                  %
%			S i m u l a t i o n   T e c h n i q u e s                                              %
%                                                                                                  %
%--------------------------------------------------------------------------------------------------%
\subsection{Simulation Techniques}
%--------------------------------------------------------------------------------------------------%
\label{subsec: Simulation}
%--------------------------------------------------------------------------------------------------%

All circuits are simulated exactly and without noise, with outcome probabilities computed rather than sampled. Three observations make full-scale training affordable.

\noindent\textbf{Amplitude encoding.}
One matrix product per objective evaluation, where the ansatz consists of $R_{y}$ rotations and CX gates, so $U(\boldsymbol{\theta})$ is a real orthogonal matrix that depends on $\boldsymbol{\theta}$ and not on the data. Amplitude encoding maps a record directly to the unit vector $\hat{\mathbf{x}}$. The outcome probabilities of all $m$ records are therefore the squared entries of one matrix product:
%--------------------------------------------------------------------------------------------------%
\begin{align}
	P
	=
	\big( \hat{X} U^{\top}(\boldsymbol{\theta}) \big) \odot \big( \hat{X} U^{\top}(\boldsymbol{\theta}) \big)
	\ ,
	\quad P_{jb} = p_{j}(b)
	\ ,
	\label{eq: Amplitude Pass}
\end{align}
%--------------------------------------------------------------------------------------------------%
where the rows of $\hat{X}$ are the normalized records, instead of $m$ circuit simulations; on four qubits $U$ is a $16 \times 16$ matrix built once per objective evaluation.

\noindent\textbf{Angle encoding.}
Gates applied to a batch of states, where at 16 qubits a dense $U$ would need 34~GB, so the angle circuit is simulated gate by gate on a batch of state vectors, each reshaped into a tensor with one axis per qubit (Algorithm~\ref{alg: Angle}). Two exact reductions remove most of the work.
The encoding rotation $R_{y}(s x_{i})$ and the first trainable rotation $R_{y}(\theta_{i})$ act on the same qubit with nothing in between, and $R_{y}(\theta) R_{y}(\alpha) = R_{y}(\theta + \alpha)$, so the encoding is absorbed into the first layer and the encoded state is never built separately.
A CX gate permutes basis states and is applied as an exchange of two slices, with no arithmetic.
All $\langle Z_{S} \rangle$ needed by the head are obtained at once by a fast Walsh--Hadamard transform of $p(b)$, at cost $n 2^{n}$ per record. The fidelity kernels of Equation~\eqref{eq: Fidelity Kernels} are evaluated in closed form, without state vectors, so the cost of the reference grows only linearly with $N$.

%--------------------------------------------------------------------------------------------------%
\begin{algorithm}[H]
	%----------------------------------------------------------------------------------------------%
	\small
	%----------------------------------------------------------------------------------------------%
	\caption{Angle-encoded forward pass on a batch of state vectors.}
	\label{alg: Angle}
	%----------------------------------------------------------------------------------------------%
	\DontPrintSemicolon
	%----------------------------------------------------------------------------------------------%
	\KwIn{records $X \in \mathbb{R}^{m \times n}$, scale $s$, parameters $\boldsymbol{\theta}$, repetitions $r$}
	\KwOut{$\langle Z_{S} \rangle$ for the requested strings $S$, for all $m$ records}
	$\boldsymbol{\psi} \gets \bigotimes_{i} R_{y}(s x_{i} + \theta_{i}) \ket{0}$ for every record \tcp*{encoding absorbed into layer 0}
	\For{$\ell \gets 1$ \KwTo $r$}
	{
		\For{$q \gets 0$ \KwTo $n - 2$}
		{
			swap the slices $\ket{1}_{q} \ket{0}_{q + 1}$ and $\ket{1}_{q} \ket{1}_{q + 1}$ of $\boldsymbol{\psi}$ \tcp*{CX$(q, q + 1)$ as a permutation}
		}
		apply $R_{y}(\theta_{\ell n + q})$ to axis $q$ of $\boldsymbol{\psi}$ for every $q$ \tcp*{one contraction per qubit}
	}
	$p \gets \abs{\boldsymbol{\psi}}^{2}$; \quad $\hat{p} \gets \mathrm{FWHT}(p)$ \tcp*{$\hat{p}_{S} = \langle Z_{S} \rangle$ for every $S$}
	\Return $\hat{p}_{S}$ for the requested $S$
	%----------------------------------------------------------------------------------------------%
\end{algorithm}
%--------------------------------------------------------------------------------------------------%

%	\emph{Verification.}
Both simulators are checked against Qiskit: the amplitude forward pass against \texttt{SamplerQNN} of Qiskit Machine Learning \cite{Sahin2025} on five configurations (largest difference in outcome probability $7.9 \times 10^{-12}$) and the angle forward pass against \texttt{Statevector} \cite{JavadiAbhari2024} on four configurations ($8.9 \times 10^{-16}$); the dense ansatz of the amplitude arm of E4 agrees with Qiskit to $5.6 \times 10^{-16}$. A training run of the published four-qubit circuit on 20,153 records takes a median of 1.2~s, against $15.6 \pm 2.1$~min reported for 2,000 records in \cite{Thomos2026} on the same CPU model (not a controlled timing).
%	Table~S12 maps each technique to the file and function that implements it, so that every number can be traced to the code.
Analyses use NumPy \cite{Harris2020}, SciPy \cite{Virtanen2020}, scikit-learn \cite{Pedregosa2011}, Qiskit \cite{JavadiAbhari2024} and Qiskit Machine Learning \cite{Sahin2025}.

\subsection{Experimental Design \& Statistics}
%--------------------------------------------------------------------------------------------------%
\label{subsec: Design}
%--------------------------------------------------------------------------------------------------%

Each experiment varies one factor and holds the others fixed; encoding-level metrics use the encoded states alone, classifier-level metrics a trained model.
E1 applies every readout to one frozen anchor circuit (amplitude, $N = 16$, 20,153 records), adds a co-trained head, and computes the attainable reference on the same 20,153 records.
E2 measures separability and the reference of all three encodings at $N = 4$, 8 and 16 (2,000 records).
E3 removes the norm or restores it on a fifth qubit, with a depth-matched control.
E4 is the matched comparison of the two encodings with three readouts and their references (2,000 records; the 16-qubit angle circuit at seed 0 only).
E5 crosses depth ($\textit{reps} \in \{1, 3, 5, 9, 17\}$) with the readout at the protocol of \cite{Thomos2026}.
E6 reads the seed-0 angle circuits of E4 with finite shots, and E7 transpiles the circuits of E1 and E4.
All use five seeds unless stated.

\emph{Anchor.}
The anchor (E1, qubit-0 parity) is the layout of the control study under the new protocol, and E5 repeats the published depth sweep with the same subsample, budget schedule and initialization.
Published numbers are compared only there (Section~\ref{subsec: Anchor}); every other comparison is made within the new protocol.

\emph{Shots and resources.}
In E6 outcomes are sampled from the exact distribution (parity: $10^{2}$ to $10^{5}$ shots, 20 draws; head: $10^{3}$ and $10^{4}$ shots, 5 draws, all features from the same samples).
In E7 circuits are transpiled with Qiskit 2.3.0 (optimization level 3) to $\{\mathrm{CX}, U\}$ with all-to-all or linear-chain connectivity, amplitude state preparation by \texttt{StatePreparation} on 25 records per $N$.

\emph{Statistics.}
% REFS-CUT: Seeds are compared with Welch's $t$-test \cite{welch1947} for independent runs (including published means) and the paired $t$-test when a seed fixes data and initialisation for both arms, as in E5, with Hedges' $g$ \cite{hedges1981} or $d_{z}$ as effect sizes and Student-$t$ intervals over seeds; a record-level bootstrap is used only for single fixed models.
Seeds are compared with Welch's $t$-test for independent runs (including published means) and the paired $t$-test when a seed fixes data and initialization for both arms, as in E4, with Hedges' $g$ or $d_{z}$ as effect sizes and Student-$t$ intervals over seeds; a record-level bootstrap is used only for single fixed models.
$p$-values are descriptive and uncorrected for multiple comparisons.
With five seeds a non-significant difference is not evidence of equality: at the anchor's dispersion the 95\% interval of a five-seed mean spans 7 points of internal accuracy and 22 of novel recall, and the interval of a difference between two such means is wider still.
\section{Results}
%--------------------------------------------------------------------------------------------------%
\label{sec: Results}
\subsection{The Control Study under the New Protocol}
%--------------------------------------------------------------------------------------------------%
\label{subsec: Anchor}
%--------------------------------------------------------------------------------------------------%

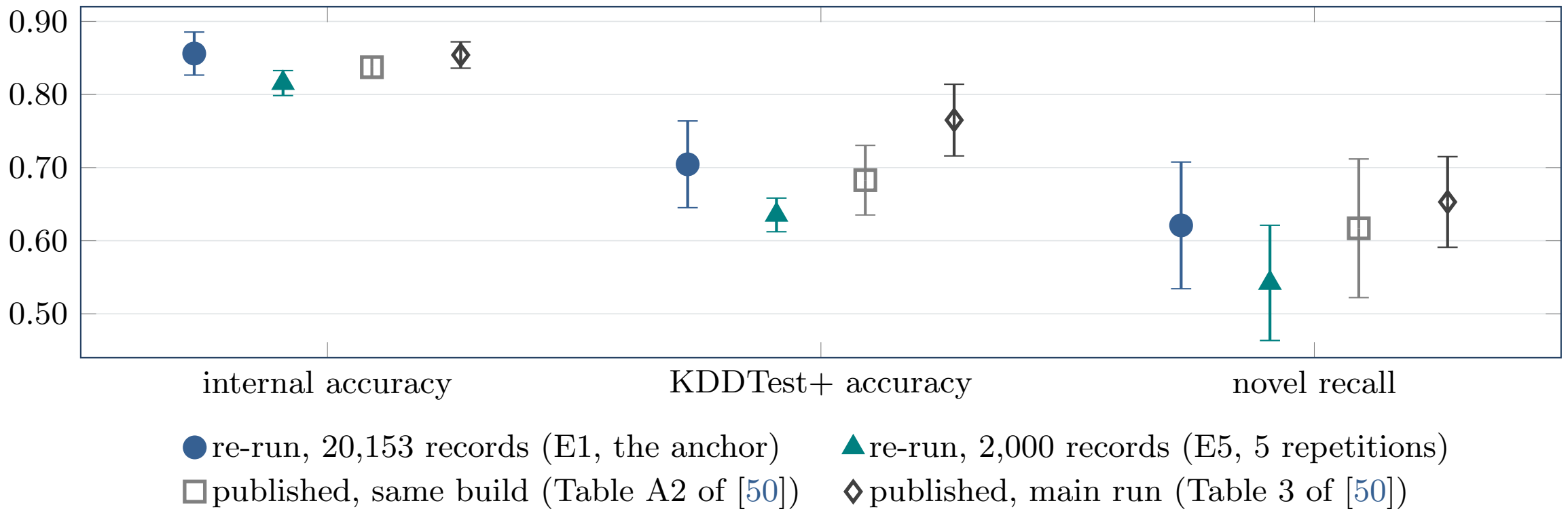
\begin{figure}[H]
	%----------------------------------------------------------------------------------------------%
	\centering
	%----------------------------------------------------------------------------------------------%
	\begin{tikzpicture}
		%------------------------------------------------------------------------------------------%
		\begin{axis}
			[
				width = \linewidth, height = 5.000 cm,
				xmin = 0.500, xmax = 3.500, xtick = { 1, 2, 3 },
				xticklabels = { internal accuracy, KDDTest+ accuracy, novel recall },
				ymin = 0.440, ymax = 0.920, ytick = { 0.50, 0.60, 0.70, 0.80, 0.90 },
				yticklabel style = { /pgf/number format/fixed, /pgf/number format/precision = 2, /pgf/number format/zerofill },
				ymajorgrids = true, grid style = { WordIceBlue, line width = 0.300 pt },
				axis line style = { WordBlueDarker50 }, tick label style = { font = \small },
				label style = { font = \small },
				error bars/y dir = both, error bars/y explicit,
				error bars/error bar style = { line width = 0.800 pt },
				every axis plot/.append style = { only marks, mark size = 2.800 pt, line width = 1.000 pt },
				legend style = { font = \small, draw = none, fill = none, cells = { anchor = west }, at = { (0.500, -0.170) }, anchor = north, /tikz/every even column/.append style = { column sep = 10pt } },
				legend columns = 2,
			]
			\addplot [ WordBlueDarker25, mark = * ] coordinates { (0.73, 0.8560) +- (0, 0.0294) (1.73, 0.7045) +- (0, 0.0593) (2.73, 0.6210) +- (0, 0.0866) };
			\addlegendentry{re-run, 20,153 records (E1, the anchor)}
			\addplot [ GreenTeal, mark = triangle* ] coordinates { (0.91, 0.8156) +- (0, 0.0171) (1.91, 0.6353) +- (0, 0.0230) (2.91, 0.5423) +- (0, 0.0788) };
			\addlegendentry{re-run, 2,000 records (E5, 5 repetitions)}
			\addplot [ gray, mark = square ] coordinates { (1.09, 0.8374) +- (0, 0.0133) (2.09, 0.6828) +- (0, 0.0476) (3.09, 0.6170) +- (0, 0.0948) };
			\addlegendentry{published, same build (Table A2 of \cite{Thomos2026})}
			\addplot [ black!75, mark = diamond ] coordinates { (1.27, 0.8540) +- (0, 0.0180) (2.27, 0.7650) +- (0, 0.0490) (3.27, 0.6530) +- (0, 0.0620) };
			\addlegendentry{published, main run (Table 3 of \cite{Thomos2026})}
		\end{axis}
		%------------------------------------------------------------------------------------------%
	\end{tikzpicture}
	%----------------------------------------------------------------------------------------------%
	\caption{The control study under the new protocol: the published four-qubit circuit (amplitude encoding of 16 components, five repetitions, qubit-0 parity) re-run on all 20,153 training records (E1) and on the 2,000-record subsample of the published protocol (E5), against the published runs on the same build (Table A2 of \cite{Thomos2026}, five repetitions) and on the original build (Table 3 of \cite{Thomos2026}).
	Mean $\pm$ SD over five seeds; recall on the 3,752 novel records.}
	\label{fig: Anchor}
	%----------------------------------------------------------------------------------------------%
\end{figure}

Re-run with the preprocessing of Section~\ref{subsec: Data} and the exact simulator, the published four-qubit circuit (the anchor), trained on all 20,153 records, gives $0.856 \pm 0.029$ internal accuracy, $0.704 \pm 0.059$ on KDDTest+ and $0.621 \pm 0.087$ novel recall over five seeds: within $+1.9$, $+2.2$ and $+0.4$ points of the published run on the same build (Table A2 of \cite{Thomos2026}; Welch $p \geq 0.25$; Figure~\ref{fig: Anchor}). Against the published main run (Table 3 of \cite{Thomos2026}) internal accuracy and novel recall agree ($p \geq 0.52$) and KDDTest+ accuracy is 6.1 points lower ($p = 0.12$), the size of the 8.2-point difference that the control study reports between its own two builds. The repeated depth sweep, on the 2,000-record subsample of the published protocol, agrees with the published one in internal accuracy within 2.2 points at every depth
%	($p \geq 0.056$; Table~S7).
%	TONE-CHECK: The re-run therefore reproduces the published numbers within their seed-to-seed dispersion.
The re-run therefore agrees with the published numbers within their seed-to-seed dispersion.
Two differences do matter: training on all 20,153 records instead of 2,000 adds 4.0 points to the anchor (95\% CI $+0.4$ to $+7.7$), and logistic regression on the normalised components gains 2.0 to 2.4 points of internal accuracy over the published baseline (0.951 against 0.927) at unchanged KDDTest+ accuracy and novel recall, consistent with the preprocessing of Section~\ref{subsec: Data}.

%--------------------------------------------------------------------------------------------------%
\subsection{How Much Class Signal Do the Encoded States Retain?}
%	\subsection{RQ1: How Much Class Signal Do the Encoded States Retain?}
%--------------------------------------------------------------------------------------------------%
\label{subsec: RQ1}
%--------------------------------------------------------------------------------------------------%

\begin{figure}[H]
	%----------------------------------------------------------------------------------------------%
	\centering
	%----------------------------------------------------------------------------------------------%
	\begin{tikzpicture}
		%------------------------------------------------------------------------------------------%
		\begin{axis}
			[
				name = GeometryA,
				width = 0.500\linewidth, height = 4.600 cm,
				xmode = log, log basis x = 2, xmin = 3.3, xmax = 19.5,
				xtick = { 4, 8, 16 }, xticklabels = { 4, 8, 16 },
				yticklabel style = { /pgf/number format/fixed, /pgf/number format/precision = 2, /pgf/number format/zerofill },
				xlabel = { principal components $N$ },
				grid = major, grid style = { WordIceBlue, line width = 0.300 pt },
				axis line style = { WordBlueDarker50 }, tick label style = { font = \footnotesize },
				label style = { font = \footnotesize }, title style = { font = \footnotesize\bfseries },
				error bars/y dir = both, error bars/y explicit,
				error bars/error bar style = { line width = 0.500 pt },
				every axis plot/.append style = { line width = 1.100 pt, mark size = 2.200 pt },
				legend style = { font = \scriptsize, draw = none, fill = white, fill opacity = 0.85, text opacity = 1, cells = { anchor = west } },
				title = {(a) Pairwise separability}, ylabel = { AUC }, ymin = 0.500, ymax = 0.860, ytick = { 0.50, 0.60, 0.70, 0.80 }, legend to name = GeometryLegend, legend columns = 3,
			]
			\addplot [ WordBlueDarker25, mark = * ] coordinates { (4, 0.6105) +- (0, 0.0075) (8, 0.5580) +- (0, 0.0047) (16, 0.5470) +- (0, 0.0048) };
			\addlegendentry{amplitude ($\log_{2} N$ qubits)}
			\addplot [ MagentaDark, mark = square* ] coordinates { (4, 0.8019) +- (0, 0.0071) (8, 0.7598) +- (0, 0.0039) (16, 0.7550) +- (0, 0.0042) };
			\addlegendentry{angle-RY ($N$ qubits)}
			\addplot [ GreenTeal, mark = triangle* ] coordinates { (4, 0.8278) +- (0, 0.0084) (8, 0.7953) +- (0, 0.0069) (16, 0.7837) +- (0, 0.0084) };
			\addlegendentry{angle-phase ($N$ qubits)}
		\end{axis}
		%------------------------------------------------------------------------------------------%
		\begin{axis}
			[
				at = { ($ (GeometryA.east) + (1.600 cm, 0) $) }, anchor = west,
				width = 0.500\linewidth, height = 4.600 cm,
				xmode = log, log basis x = 2, xmin = 3.3, xmax = 19.5,
				xtick = { 4, 8, 16 }, xticklabels = { 4, 8, 16 },
				yticklabel style = { /pgf/number format/fixed, /pgf/number format/precision = 2, /pgf/number format/zerofill },
				xlabel = { principal components $N$ },
				grid = major, grid style = { WordIceBlue, line width = 0.300 pt },
				axis line style = { WordBlueDarker50 }, tick label style = { font = \footnotesize },
				label style = { font = \footnotesize }, title style = { font = \footnotesize\bfseries },
				error bars/y dir = both, error bars/y explicit,
				error bars/error bar style = { line width = 0.500 pt },
				every axis plot/.append style = { line width = 1.100 pt, mark size = 2.200 pt },
				title = {(b) Attainable reference ($C = 1$)}, ylabel = { internal accuracy }, ymin = 0.880, ymax = 0.975, ytick = { 0.88, 0.90, 0.92, 0.94, 0.96 },
			]
			\addplot [ WordBlueDarker25, mark = * ] coordinates { (4, 0.8939) +- (0, 0.0029) (8, 0.9422) +- (0, 0.0052) (16, 0.9606) +- (0, 0.0033) };
			\addplot [ MagentaDark, mark = square* ] coordinates { (4, 0.9312) +- (0, 0.0043) (8, 0.9541) +- (0, 0.0044) (16, 0.9573) +- (0, 0.0039) };
			\addplot [ GreenTeal, mark = triangle* ] coordinates { (4, 0.9328) +- (0, 0.0041) (8, 0.9656) +- (0, 0.0042) (16, 0.9670) +- (0, 0.0040) };
		\end{axis}
		%------------------------------------------------------------------------------------------%
	\end{tikzpicture}
	%----------------------------------------------------------------------------------------------%
	\\[0.150 cm]
	\pgfplotslegendfromname{GeometryLegend}
	%----------------------------------------------------------------------------------------------%
	\caption{What the encoded states retain (E2): (a) separability AUC of Equation~\eqref{eq: AUC} and (b) attainable reference with $C = 1$, for the three encodings on the same 2,000 training records per seed; the reference is scored on 2,000 records of the internal partition.
	Mean $\pm$ SD over five seeds; all values, including $C = 100$.
	%	, in Table~S2.
	}
	\label{fig: Geometry}
	%----------------------------------------------------------------------------------------------%
\end{figure}
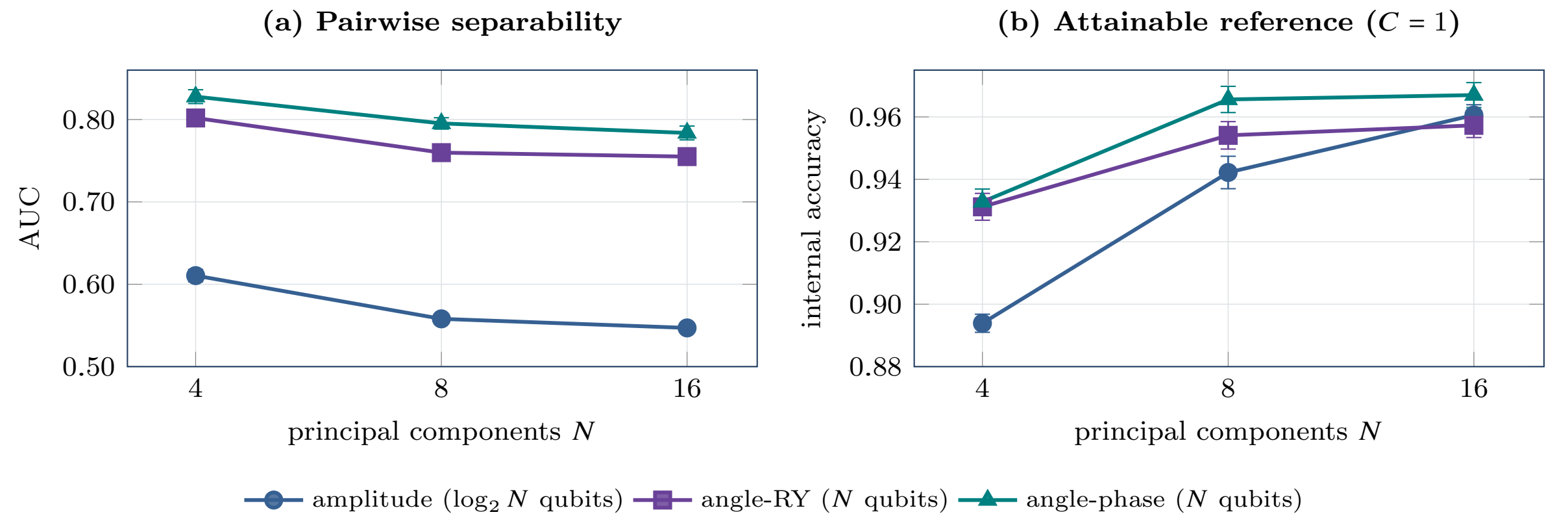

In pairwise separability angle encoding leads by about 0.2 AUC at every size (angle-RY 0.802, 0.760 and 0.755 against amplitude 0.610, 0.558 and 0.547 at $N = 4$, 8 and 16; paired $p < 0.001$; Figure~\ref{fig: Geometry}a), retaining 5.4 to 6.0 times the above-chance signal of amplitude encoding at $N = 16$.
On the scale of accuracy the advantage is much smaller and shrinks with $N$: the attainable reference of angle-RY encoding exceeds that of amplitude encoding by 3.7 and 1.2 points at $N = 4$ and 8 and is 0.3 points below it at $N = 16$ ($C = 1$; 4.0, 1.0 and 0.7 points with $C = 100$; Figure~\ref{fig: Geometry}b).
From 4 to 16 components the two measures move in opposite directions: the amplitude AUC falls by 0.063 while its reference rises by 6.7 points, because the AUC weighs all pairs equally whereas accuracy depends on the records near the decision boundary.
Trained on all 20,153 records, the reference of the amplitude-encoded states at $N = 16$ reaches 0.970 ($C = 1$) and 0.984 ($C = 100$), above logistic regression on the same input and records (0.951; E1).
The angle-RY kernel is practically a Gaussian kernel with $\gamma = s^{2} / 4$ (accuracies within 0.4 points, kernel matrices correlated above 0.99).
%	; Table~S5).
%--------------------------------------------------------------------------------------------------%
% TONE-CHECK: The norm that amplitude encoding removes is informative on its own (AUC 0.870), but restoring it helps neither logistic regression (0.949 against 0.951) nor the circuit ($-1.2$ points against a depth-matched control, 95\% CI $-7.6$ to $+5.3$; Table~S3).
%--------------------------------------------------------------------------------------------------%
The norm that amplitude encoding removes is informative on its own (AUC 0.870), but restoring it gives no measurable gain to logistic regression (0.949 against 0.951) or to the circuit ($-1.2$ points against a depth-matched control, 95\% CI $-7.6$ to $+5.3$.
%	; Table~S3
).

%--------------------------------------------------------------------------------------------------%
\subsection{How Close Do Trained Circuits Come to Their Reference?}
%	\subsection{RQ2: How Close Do Trained Circuits Come to Their Reference?}
%--------------------------------------------------------------------------------------------------%
\label{subsec: RQ2}
%--------------------------------------------------------------------------------------------------%

\begin{figure}[H]
	%----------------------------------------------------------------------------------------------%
	\centering
	%----------------------------------------------------------------------------------------------%
	\begin{tikzpicture}
		%------------------------------------------------------------------------------------------%
		\begin{axis}
			[
				width = 0.820\linewidth, height = 5.400 cm,
				xmin = 0.400, xmax = 1.000, xtick = { 0.40, 0.50, 0.60, 0.70, 0.80, 0.90, 1.00 },
				xticklabel style = { /pgf/number format/fixed, /pgf/number format/precision = 2, /pgf/number format/zerofill },
				xlabel = { internal accuracy },
				ymin = 0.400, ymax = 6.600, ytick = { 1, 2, 3, 4, 5, 6 },
				yticklabels = { global parity, qubit-0 parity, best fixed assignment, frozen linear head, co-trained linear head, MLP head },
				xmajorgrids = true, grid style = { WordIceBlue, line width = 0.300 pt },
				axis line style = { WordBlueDarker50 }, axis x line* = bottom, axis y line* = left,
				tick label style = { font = \small }, label style = { font = \small },
				legend style = { font = \small, draw = none, fill = none, cells = { anchor = west }, at = { (0.500, -0.280) }, anchor = north, /tikz/every even column/.append style = { column sep = 8pt } },
				legend columns = 2,
			]
			\addplot [ forget plot, xbar, bar width = 10.000 pt, fill = gray!45, draw = none, error bars/x dir = both, error bars/x explicit, error bars/error bar style = { line width = 0.800 pt, black!70 } ] coordinates { (0.5400, 1) +- (0.1354, 0) };
			\addplot [ forget plot, xbar, bar width = 10.000 pt, fill = WordBlueDarker25, draw = none, error bars/x dir = both, error bars/x explicit, error bars/error bar style = { line width = 0.800 pt, black!70 } ] coordinates { (0.8560, 2) +- (0.0294, 0) (0.8921, 3) +- (0.0090, 0) (0.9078, 4) +- (0.0071, 0) (0.9481, 5) +- (0.0068, 0) };
			\addplot [ forget plot, xbar, bar width = 10.000 pt, fill = GreenTeal, draw = none, error bars/x dir = both, error bars/x explicit, error bars/error bar style = { line width = 0.800 pt, black!70 } ] coordinates { (0.9829, 6) +- (0.0015, 0) };
			\node [ font = \footnotesize, anchor = west ] at (axis cs: 0.6850, 1) { 0.540 };
			\node [ font = \footnotesize, text = white, anchor = west ] at (axis cs: 0.4080, 2) { 0.856 };
			\node [ font = \footnotesize, text = white, anchor = west ] at (axis cs: 0.4080, 3) { 0.892 };
			\node [ font = \footnotesize, text = white, anchor = west ] at (axis cs: 0.4080, 4) { 0.908 };
			\node [ font = \footnotesize, text = white, anchor = west ] at (axis cs: 0.4080, 5) { 0.948 };
			\node [ font = \footnotesize, text = white, anchor = west ] at (axis cs: 0.4080, 6) { 0.983 };
			\addplot [ RedPurple, dashed, line width = 1.100 pt ] coordinates { (0.9700, 0.400) (0.9700, 6.600) };
			\addlegendentry{reference, $C = 1$ (0.970)}
			\addplot [ RedPurple, densely dotted, line width = 1.100 pt ] coordinates { (0.9835, 0.400) (0.9835, 6.600) };
			\addlegendentry{reference, $C = 100$ (0.984)}
			\addplot [ gray, densely dotted, line width = 1.000 pt ] coordinates { (0.9512, 0.400) (0.9512, 6.600) };
			\addlegendentry{logistic regression on $\hat{\mathbf{x}}$ (0.951)}
		\end{axis}
		%------------------------------------------------------------------------------------------%
	\end{tikzpicture}
	%----------------------------------------------------------------------------------------------%
	\caption{Readouts of the anchor circuit (E1): amplitude encoding of 16 components on four qubits, five repetitions, 20,153 training records.
	Every readout except the co-trained head uses the weights trained for qubit-0 parity; global parity (grey) is applied to the same weights, the MLP head (green) is not linear.
	Mean $\pm$ SD over five seeds.
	Vertical lines: attainable reference of the same states on the same records and logistic regression on the same input.
	%	; values in Table~S1.
	}
	\label{fig: Readouts}
	%----------------------------------------------------------------------------------------------%
\end{figure}
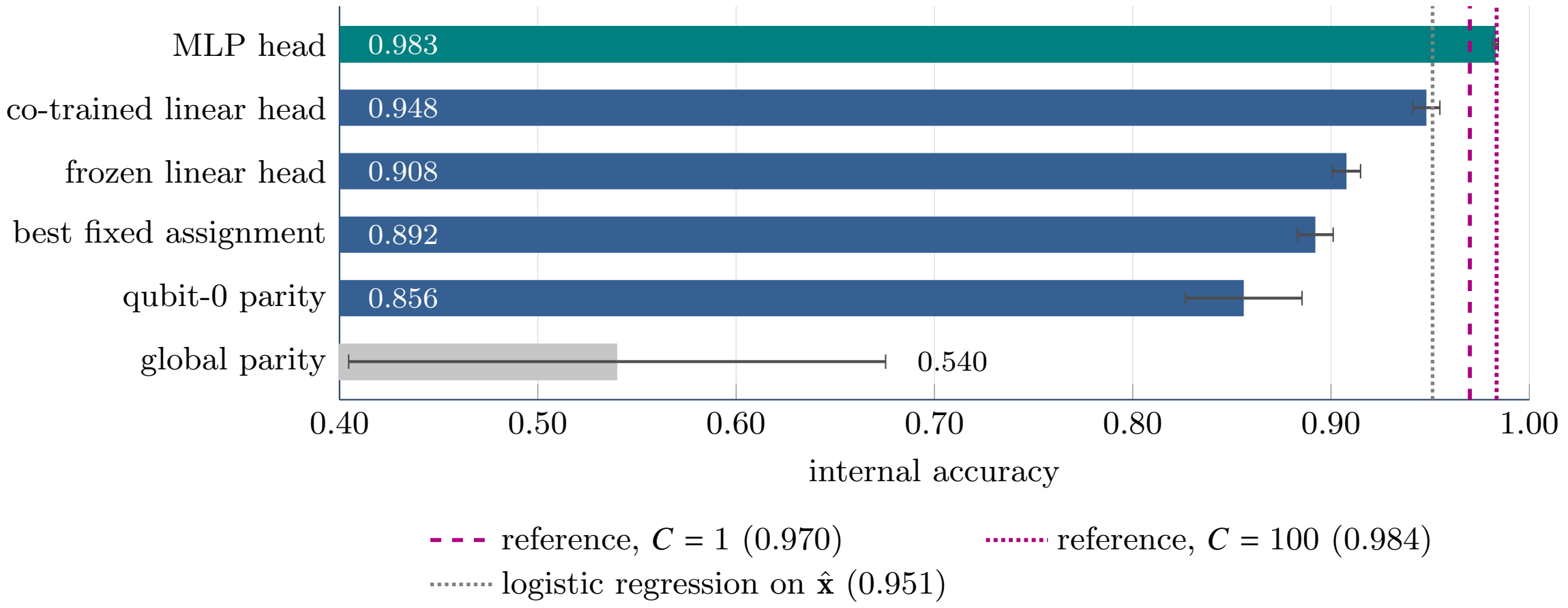

\noindent\textbf{Readout of the published circuit.}
%--------------------------------------------------------------------------------------------------%
% TONE-CHECK: On the anchor circuit (amplitude, $N = 16$), every readout other than qubit-0 parity raises internal accuracy (E1; Figure~\ref{fig: Readouts}, Table~S1): the best fixed assignment by 3.6 points (0.892), a frozen linear head by 5.2 points (0.908), a co-trained linear head by 9.2 points (0.948), and a non-linear MLP head by 12.7 points (0.983).
%--------------------------------------------------------------------------------------------------%
On the anchor circuit (amplitude, $N = 16$), every readout fitted on the training records raises internal accuracy above qubit-0 parity (E1; Figure~\ref{fig: Readouts}): the best fixed assignment by 3.6 points (0.892), a frozen linear head by 5.2 points (0.908), a co-trained linear head by 9.2 points (0.948), and a non-linear MLP head by 12.7 points (0.983).
Against the attainable reference of these states on the same 20,153 training records (0.970, $C = 1$), qubit-0 parity is 11.4 points short, the frozen head 6.2 and the co-trained head 2.2; the MLP head exceeds the reference by 1.3 points, as a non-linear readout can (Section~\ref{subsec: Reference}), and is within 0.1 point of the reference with $C = 100$ (0.984).
Global parity, applied to the same circuit trained for qubit-0 parity, gives 0.540, so the two parity conventions are not interchangeable.

%--------------------------------------------------------------------------------------------------%
%--------------------------------------------------------------------------------------------------%

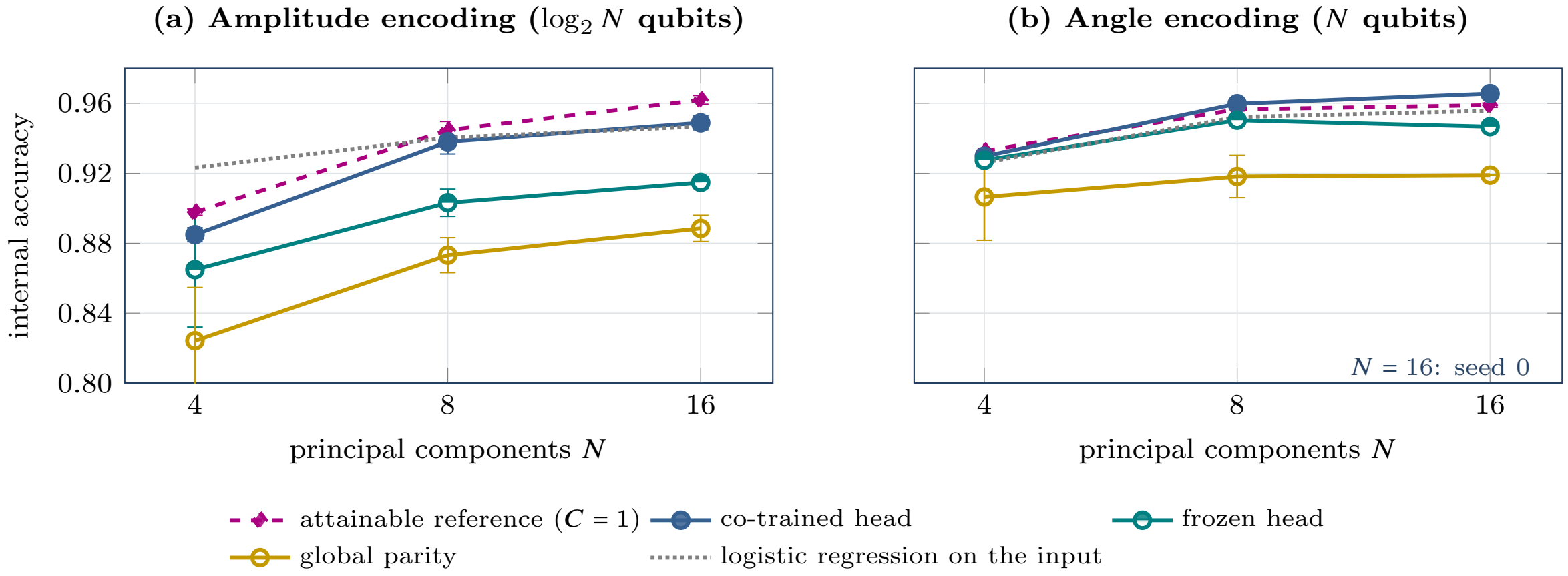
\begin{figure}[H]
	%----------------------------------------------------------------------------------------------%
	\centering
	%----------------------------------------------------------------------------------------------%
	\begin{tikzpicture}
		%------------------------------------------------------------------------------------------%
		\begin{axis}
			[
				name = MatchedA,
				width = 0.500\linewidth, height = 4.700 cm,
				xmode = log, log basis x = 2, xmin = 3.3, xmax = 19.5,
				xtick = { 4, 8, 16 }, xticklabels = { 4, 8, 16 },
				ymin = 0.800, ymax = 0.980, ytick = { 0.80, 0.84, 0.88, 0.92, 0.96 },
				yticklabel style = { /pgf/number format/fixed, /pgf/number format/precision = 2, /pgf/number format/zerofill },
				xlabel = { principal components $N$ }, ylabel = { internal accuracy },
				grid = major, grid style = { WordIceBlue, line width = 0.300 pt },
				axis line style = { WordBlueDarker50 }, tick label style = { font = \footnotesize },
				label style = { font = \footnotesize }, title style = { font = \footnotesize\bfseries },
				error bars/y dir = both, error bars/y explicit,
				error bars/error bar style = { line width = 0.500 pt },
				every axis plot/.append style = { line width = 1.100 pt, mark size = 2.200 pt },
				legend style = { font = \scriptsize, draw = none, fill = white, fill opacity = 0.85, text opacity = 1, cells = { anchor = west } },
				title = {(a) Amplitude encoding ($\log_{2} N$ qubits)}, legend to name = MatchedLegend, legend columns = 3,
			]
			\addplot [ RedPurple, dashed, mark = diamond* ] coordinates { (4, 0.8978) +- (0, 0.0018) (8, 0.9446) +- (0, 0.0050) (16, 0.9619) +- (0, 0.0025) };
			\addlegendentry{attainable reference ($C = 1$)}
			\addplot [ WordBlueDarker25, mark = * ] coordinates { (4, 0.8849) +- (0, 0.0040) (8, 0.9380) +- (0, 0.0069) (16, 0.9488) +- (0, 0.0040) };
			\addlegendentry{co-trained head}
			\addplot [ GreenTeal, mark = halfcircle* ] coordinates { (4, 0.8649) +- (0, 0.0329) (8, 0.9032) +- (0, 0.0078) (16, 0.9148) +- (0, 0.0034) };
			\addlegendentry{frozen head}
			\addplot [ WordGoldAccent1Darker25, mark = o ] coordinates { (4, 0.8242) +- (0, 0.0305) (8, 0.8732) +- (0, 0.0100) (16, 0.8885) +- (0, 0.0075) };
			\addlegendentry{global parity}
			\addplot [ gray, densely dotted, mark = none ] coordinates { (4, 0.9233) (8, 0.9403) (16, 0.9467) };
			\addlegendentry{logistic regression on the input}
		\end{axis}
		%------------------------------------------------------------------------------------------%
		\begin{axis}
			[
				at = { ($ (MatchedA.east) + (1.400 cm, 0) $) }, anchor = west,
				width = 0.500\linewidth, height = 4.700 cm,
				xmode = log, log basis x = 2, xmin = 3.3, xmax = 19.5,
				xtick = { 4, 8, 16 }, xticklabels = { 4, 8, 16 },
				ymin = 0.800, ymax = 0.980, ytick = { 0.80, 0.84, 0.88, 0.92, 0.96 }, yticklabels = {},
				xlabel = { principal components $N$ },
				grid = major, grid style = { WordIceBlue, line width = 0.300 pt },
				axis line style = { WordBlueDarker50 }, tick label style = { font = \footnotesize },
				label style = { font = \footnotesize }, title style = { font = \footnotesize\bfseries },
				error bars/y dir = both, error bars/y explicit,
				error bars/error bar style = { line width = 0.500 pt },
				every axis plot/.append style = { line width = 1.100 pt, mark size = 2.200 pt },
				title = {(b) Angle encoding ($N$ qubits)},
			]
			\addplot [ RedPurple, dashed, mark = diamond* ] coordinates { (4, 0.9328) +- (0, 0.0004) (8, 0.9565) +- (0, 0.0009) (16, 0.9589) +- (0, 0.0012) };
			\addplot [ WordBlueDarker25, mark = * ] coordinates { (4, 0.9298) +- (0, 0.0021) (8, 0.9597) +- (0, 0.0015) (16, 0.9655) +- (0, 0) };
			\addplot [ GreenTeal, mark = halfcircle* ] coordinates { (4, 0.9275) +- (0, 0.0035) (8, 0.9503) +- (0, 0.0030) (16, 0.9466) +- (0, 0) };
			\addplot [ WordGoldAccent1Darker25, mark = o ] coordinates { (4, 0.9065) +- (0, 0.0248) (8, 0.9182) +- (0, 0.0121) (16, 0.9190) +- (0, 0) };
			\addplot [ gray, densely dotted, mark = none ] coordinates { (4, 0.9262) (8, 0.9521) (16, 0.9557) };
			\node [ font = \scriptsize, text = WordBlueDarker50, anchor = north east ] at (rel axis cs: 0.97, 0.12) { $N = 16$: seed 0 };
		\end{axis}
		%------------------------------------------------------------------------------------------%
	\end{tikzpicture}
	%----------------------------------------------------------------------------------------------%
	\\[0.150 cm]
	\pgfplotslegendfromname{MatchedLegend}
	%----------------------------------------------------------------------------------------------%
	\caption{Trained circuits against the attainable reference of their own encoding (E4).
	Both encodings use $R_{y}$ layers with a linear CX chain and $3N$ parameters, COBYLA with 400 evaluations, and the same 2,000 training records and initial parameters per seed; accuracy is measured on the full internal partition.
	Mean $\pm$ SD over five seeds; the 16-qubit angle-encoded circuits are seed 0 and have no bar.
	The frozen head is refitted on the weights of the global-parity fit.
	Qubit-0 parity is omitted because of the light cone of the angle circuit (Section~\ref{subsec: Circuits}).
	%; all values are in Table~S11.
	}
	\label{fig: Matched}
	%----------------------------------------------------------------------------------------------%
\end{figure}

\noindent\textbf{Matched comparison.}
Under one protocol (E4), angle-encoded circuits are more accurate than amplitude-encoded ones at every $N$ and with every readout that reads all qubits (Figure~\ref{fig: Matched}).
With the co-trained head the advantage is 4.5 points (95\% CI 4.1 to 4.9) at $N = 4$, 2.2 points (1.3 to 3.1) at $N = 8$ and 1.6 points (seed 0) at $N = 16$; with global parity it is 8.2, 4.5 and 3.9 points.
The references of the two encodings differ by only 3.5, 1.2 and $-0.1$ points on the same records ($N = 16$: seed 0), so at $N = 16$ the trained advantage exceeds the geometric one.
The reason is the distance of each circuit from its own reference.
With global parity, amplitude-encoded circuits stay 7.4, 7.1 and 7.3 points below their reference, angle-encoded circuits 2.6, 3.8 and 4.0 points.
A frozen head closes 36\% to 55\% of that distance for amplitude encoding and 69\% to 84\% for angle encoding.
%--------------------------------------------------------------------------------------------------%
% TONE-CHECK: The co-trained head closes 82\% to 91\% of it for amplitude encoding, leaving 1.3, 0.7 and 1.3 points, and for angle encoding 89\% at $N = 4$ and all of it at $N = 8$ and 16: the angle-encoded circuit with a co-trained head reaches 0.930, 0.960 and 0.965, against references of 0.933, 0.956 and 0.959.
%--------------------------------------------------------------------------------------------------%
The co-trained head closes 82\% to 91\% of it for amplitude encoding, leaving 1.3, 0.7 and 1.3 points, and for angle encoding 89\% at $N = 4$ and all of it at $N = 8$ and 16: the angle-encoded circuit with a co-trained head reaches 0.930, 0.960 and 0.965, against references ($C = 1$) of 0.933, 0.956 and 0.959.
%--------------------------------------------------------------------------------------------------%
% TONE-CHECK: A trained angle-encoded circuit with a linear readout thus realises what its encoding allows; an amplitude-encoded circuit falls short, mostly at the readout.
%--------------------------------------------------------------------------------------------------%
A trained angle-encoded circuit with a linear readout thus comes close to what its encoding allows; an amplitude-encoded circuit falls short, mostly at the readout.
Logistic regression on the same inputs reaches 0.926, 0.952 and 0.956; the angle-encoded circuits with a co-trained head exceed it by 0.4 points (95\% CI 0.1 to 0.6, paired $p = 0.013$) at $N = 4$, by 0.8 points (0.4 to 1.1, $p = 0.003$) at $N = 8$ and by 1.0 point at $N = 16$ (seed 0), and stay 0.5, 1.9 and 1.6 points below the SVM with $C = 100$.

%--------------------------------------------------------------------------------------------------%
%--------------------------------------------------------------------------------------------------%

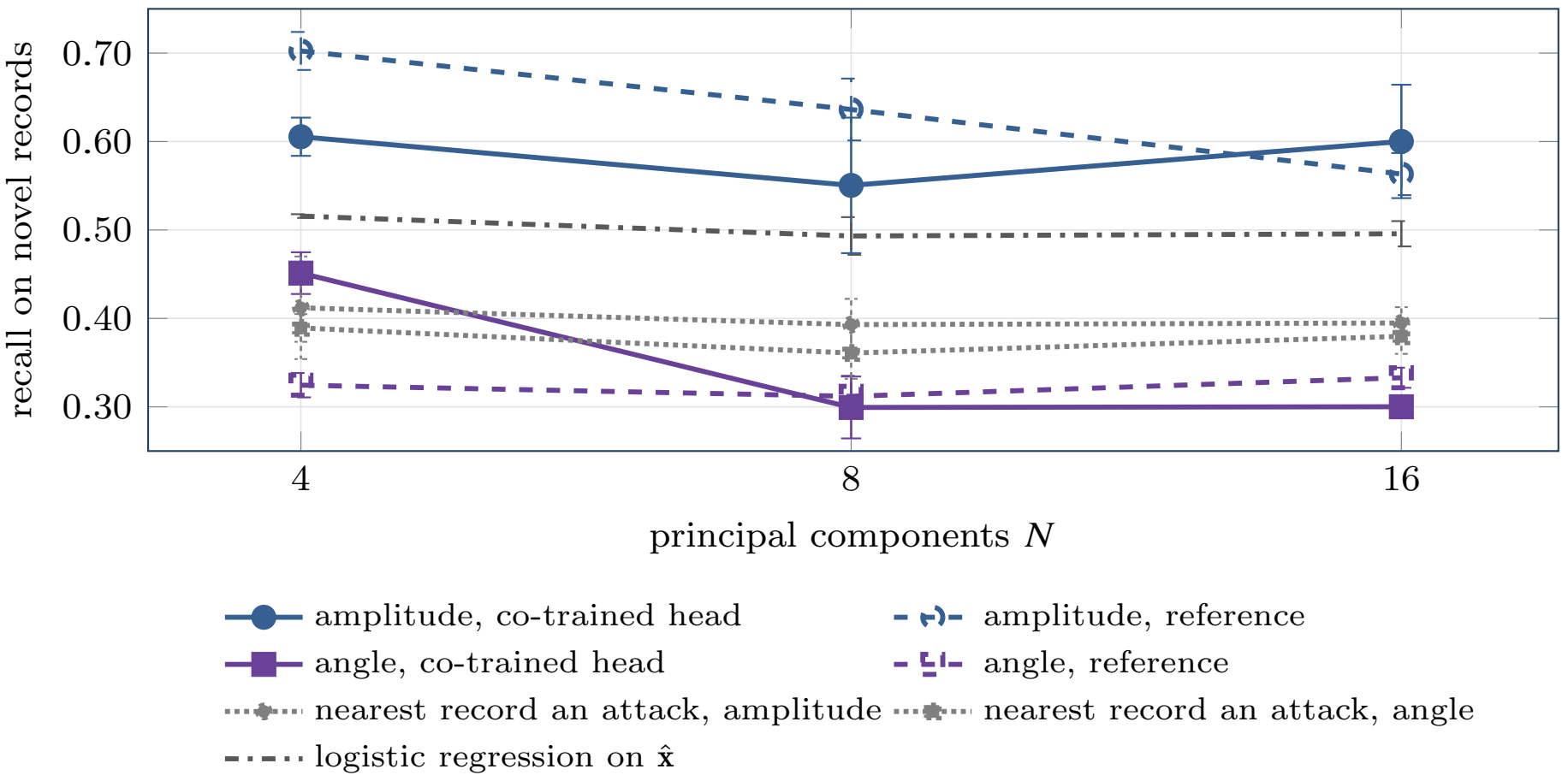
\begin{figure}[H]
	%----------------------------------------------------------------------------------------------%
	\centering
	%----------------------------------------------------------------------------------------------%
	\begin{tikzpicture}
		\begin{axis}
			[
				width = 0.780\linewidth, height = 5.000 cm,
				xmode = log, log basis x = 2, xmin = 3.3, xmax = 19.5,
				xtick = { 4, 8, 16 }, xticklabels = { 4, 8, 16 },
				ymin = 0.250, ymax = 0.750, ytick = { 0.30, 0.40, 0.50, 0.60, 0.70 },
				yticklabel style = { /pgf/number format/fixed, /pgf/number format/precision = 2, /pgf/number format/zerofill },
				xlabel = { principal components $N$ }, ylabel = { recall on novel records },
				grid = major, grid style = { WordIceBlue, line width = 0.300 pt },
				axis line style = { WordBlueDarker50 }, tick label style = { font = \footnotesize },
				label style = { font = \footnotesize },
				error bars/y dir = both, error bars/y explicit,
				error bars/error bar style = { line width = 0.500 pt },
				every axis plot/.append style = { line width = 1.100 pt, mark size = 2.200 pt },
				legend style = { font = \scriptsize, draw = none, fill = none, cells = { anchor = west }, at = { (0.500, -0.300) }, anchor = north },
				legend columns = 2,
			]
			\addplot [ WordBlueDarker25, mark = * ] coordinates { (4, 0.6054) +- (0, 0.0216) (8, 0.5504) +- (0, 0.0766) (16, 0.6001) +- (0, 0.0641) };
			\addlegendentry{amplitude, co-trained head}
			\addplot [ WordBlueDarker25, dashed, mark = o ] coordinates { (4, 0.7024) +- (0, 0.0215) (8, 0.6362) +- (0, 0.0350) (16, 0.5632) +- (0, 0.0238) };
			\addlegendentry{amplitude, reference}
			\addplot [ MagentaDark, mark = square* ] coordinates { (4, 0.4512) +- (0, 0.0236) (8, 0.2993) +- (0, 0.0350) (16, 0.3001) +- (0, 0) };
			\addlegendentry{angle, co-trained head}
			\addplot [ MagentaDark, dashed, mark = square ] coordinates { (4, 0.3245) +- (0, 0.0138) (8, 0.3119) +- (0, 0.0232) (16, 0.3329) +- (0, 0.0114) };
			\addlegendentry{angle, reference}
			\addplot [ gray, densely dotted, mark = *, mark size = 1.400 pt ] coordinates { (4, 0.4120) +- (0, 0.0580) (8, 0.3930) +- (0, 0.0291) (16, 0.3948) +- (0, 0.0179) };
			\addlegendentry{nearest record an attack, amplitude}
			\addplot [ gray, densely dotted, mark = square*, mark size = 1.400 pt ] coordinates { (4, 0.3892) +- (0, 0.0157) (8, 0.3608) +- (0, 0.0291) (16, 0.3797) +- (0, 0.0198) };
			\addlegendentry{nearest record an attack, angle}
			\addplot [ black!65, dashdotted, mark = none ] coordinates { (4, 0.5157) +- (0, 0.0021) (8, 0.4932) +- (0, 0.0213) (16, 0.4957) +- (0, 0.0143) };
			\addlegendentry{logistic regression on $\hat{\mathbf{x}}$}
		\end{axis}
	\end{tikzpicture}
	%----------------------------------------------------------------------------------------------%
	\caption{Recall on the 3,752 novel attack records (E4): circuits with a co-trained head (filled marks), the attainable references of their encodings ($C = 1$, open marks), logistic regression on the input of the amplitude-encoded circuits (dash-dotted) and the share of novel records whose most similar training record, by the fidelity of the encoding, is an attack (grey).
	Mean $\pm$ SD over five seeds; the 16-qubit angle-encoded circuit is seed 0.
	%Values in Tables~S11 and S15.
	}
	\label{fig: Novel}
	%----------------------------------------------------------------------------------------------%
\end{figure}

\noindent\textbf{Out of distribution.}
The ordering reverses out of distribution (Table~\ref{tbl: OOD}, Figure~\ref{fig: Novel}).
On KDDTest+ the co-trained amplitude-encoded circuits are 3.8 to 6.2 points more accurate, and on novel signatures they recall 0.550 to 0.605 against 0.299 to 0.451 (differences of 15 to 25 points, $N = 16$ at seed 0; 95\% CI $-19.0$ to $-11.8$ at $N = 4$, $-34.7$ to $-15.5$ at $N = 8$).
The references split the same way, so the difference belongs to the encodings, not the training.
Against logistic regression on the same input, the novel-signature recall of the co-trained amplitude-encoded circuits is 9.0 and 10.4 points higher at $N = 4$ and 16 ($p < 0.001$ and $p = 0.032$; 5.7 points at $N = 8$, $p = 0.22$), and their KDDTest+ accuracy is 5.0 and 3.7 points higher at the same $N$ ($p \leq 0.005$; $-0.6$ points at $N = 8$, $p = 0.56$).
%--------------------------------------------------------------------------------------------------%
% TONE-CHECK: Distance and norm do not explain it: novel records have a median largest fidelity to a training record of 0.978 and 0.961 at $N = 8$ and 16 under angle encoding (0.932 and 0.895 under amplitude encoding), and a median norm (3.6 to 3.7) close to normal traffic (3.3), not to known attacks (6.2).
%--------------------------------------------------------------------------------------------------%
The median largest fidelity of a novel record to a training record is 0.978 and 0.961 at $N = 8$ and 16 under angle encoding (0.932 and 0.895 under amplitude encoding), and the median norm of the novel records (3.6 to 3.7) is close to that of normal traffic (3.3) rather than to that of known attacks (6.2).
For 36\% to 41\% of the novel records the most similar training record is an attack under either encoding, and the angle-encoded circuits with a co-trained head recall roughly that share (0.30 to 0.45, within 7 points of it), as a nearest-neighbour rule would, whereas the amplitude-encoded circuits with a co-trained head exceed it by 16 to 21 points.
%	(Table~S15).

%--------------------------------------------------------------------------------------------------%
\begin{table}[H]
	%----------------------------------------------------------------------------------------------%
	\footnotesize
	%----------------------------------------------------------------------------------------------%
	\caption{Internal and KDDTest+ accuracy of the matched circuits with a co-trained head, their attainable references and logistic regression on their inputs (E4; mean over five seeds; 16-qubit angle circuit: seed 0).
	Recall on the novel records is shown in Figure~\ref{fig: Novel}.
	%; all values in Table~S11.
	}
	\label{tbl: OOD}
	%----------------------------------------------------------------------------------------------%
	\centering
	%----------------------------------------------------------------------------------------------%
	\SetTblrInner { rowsep = 0.400 mm }
	%----------------------------------------------------------------------------------------------%
	\begin{tblr}
		{
			colspec =
			{
				Q [ l, m ]
				| [ 0.500 pt, WordAquaDarker25 ]
				Q [ c, m ] Q [ c, m ] Q [ c, m ]
				| [ 0.500 pt, WordAquaDarker25 ]
				Q [ c, m ] Q [ c, m ] Q [ c, m ]
			},
			row{1-2} = { bg = WordAquaDarker25, fg = white, font = \bfseries },
			hline{1, Z} = { 1.000 pt, WordAquaDarker25 },
			hline{3} = { 0.750 pt, WordAquaDarker25 },
			hline{5, 7} = { 0.250 pt, WordAquaLighter60 },
			cell{1}{2, 5} = { c = 3 }{ c },
		}
		& Internal &&& KDDTest+ && \\
		$N$ & 4 & 8 & 16 & 4 & 8 & 16 \\
		Amplitude, co-trained head & 0.885 & 0.938 & 0.949 & 0.817 & 0.790 & 0.807 \\
		Angle, co-trained head & 0.930 & 0.960 & 0.965 & 0.755 & 0.752 & 0.745 \\
		Amplitude, reference ($C = 1$) & 0.898 & 0.945 & 0.962 & 0.838 & 0.791 & 0.815 \\
		Angle, reference ($C = 1$) & 0.933 & 0.956 & 0.959 & 0.732 & 0.748 & 0.761 \\
		Logistic regression on $\hat{\mathbf{x}}$ & 0.923 & 0.940 & 0.947 & 0.767 & 0.796 & 0.770 \\
		Logistic regression on $\mathbf{x}$ & 0.926 & 0.952 & 0.956 & 0.765 & 0.761 & 0.752 \\
	\end{tblr}\par
	%----------------------------------------------------------------------------------------------%
\end{table}
%--------------------------------------------------------------------------------------------------%

%--------------------------------------------------------------------------------------------------%
%--------------------------------------------------------------------------------------------------%

\begin{figure}[H]
	%----------------------------------------------------------------------------------------------%
	\centering
	%----------------------------------------------------------------------------------------------%
	\begin{tikzpicture}
		\begin{axis}
			[
				width = 0.760\linewidth, height = 4.500 cm,
				xmode = log, xmin = 6.5, xmax = 90,
				xtick = { 8, 16, 24, 40, 72 }, xticklabels = { 8 (1), 16 (3), 24 (5), 40 (9), 72 (17) },
				ymin = 0.760, ymax = 0.975, ytick = { 0.78, 0.82, 0.86, 0.90, 0.94 },
				yticklabel style = { /pgf/number format/fixed, /pgf/number format/precision = 2, /pgf/number format/zerofill },
				xlabel = { trainable parameters (repetitions) }, ylabel = { internal accuracy },
				grid = major, grid style = { WordIceBlue, line width = 0.300 pt },
				axis line style = { WordBlueDarker50 }, tick label style = { font = \footnotesize },
				label style = { font = \footnotesize },
				error bars/y dir = both, error bars/y explicit,
				every axis plot/.append style = { line width = 1.100 pt, mark size = 2.200 pt },
				legend pos = outer north east,
				legend style = { font = \scriptsize, draw = none, cells = { anchor = west } },
			]
			\addplot [ RedPurple, dashed, mark = none, domain = 6.5:90 ] { 0.9619 };
			\addlegendentry{attainable reference}
			\addplot [ WordBlueDarker25, mark = * ] coordinates { (8, 0.9174) +- (0, 0.0062) (16, 0.9237) +- (0, 0.0079) (24, 0.9283) +- (0, 0.0046) (40, 0.9328) +- (0, 0.0083) (72, 0.9372) +- (0, 0.0067) };
			\addlegendentry{co-trained head}
			\addplot [ GreenTeal, mark = triangle* ] coordinates { (8, 0.8879) +- (0, 0.0281) (16, 0.9033) +- (0, 0.0346) (24, 0.8827) +- (0, 0.0312) (40, 0.9018) +- (0, 0.0103) (72, 0.9007) +- (0, 0.0045) };
			\addlegendentry{best fixed assignment}
			\addplot [ WordGoldAccent1Darker25, mark = o ] coordinates { (8, 0.7883) +- (0, 0.0163) (16, 0.8524) +- (0, 0.0439) (24, 0.8156) +- (0, 0.0171) (40, 0.8640) +- (0, 0.0304) (72, 0.8818) +- (0, 0.0136) };
			\addlegendentry{qubit-0 parity}
			\addplot [ gray, only marks, mark = square, mark size = 2.400 pt, line width = 0.800 pt ] coordinates { (8, 0.7824) (16, 0.8514) (24, 0.8374) (40, 0.8780) (72, 0.8794) };
			\addlegendentry{parity, published \cite{Thomos2026}}
		\end{axis}
	\end{tikzpicture}
	%----------------------------------------------------------------------------------------------%
	\caption{Depth and readout at the protocol of the control study (E5): four-qubit RealAmplitudes circuit, amplitude encoding of 16 components, 2,000 training records, COBYLA budget $\max(150, 24(\textit{reps} + 1))$.
	Mean $\pm$ SD over five seeds.
	Squares: the published depth sweep (Table A2 of \cite{Thomos2026}).
	Dashed line: attainable reference of the encoded states ($C = 1$, 0.962), trained on the same 2,000 records per seed and scored on the full internal partition (E4).
	%	Values in Table~S10.
	}
	\label{fig: Depth}
	%----------------------------------------------------------------------------------------------%
\end{figure}
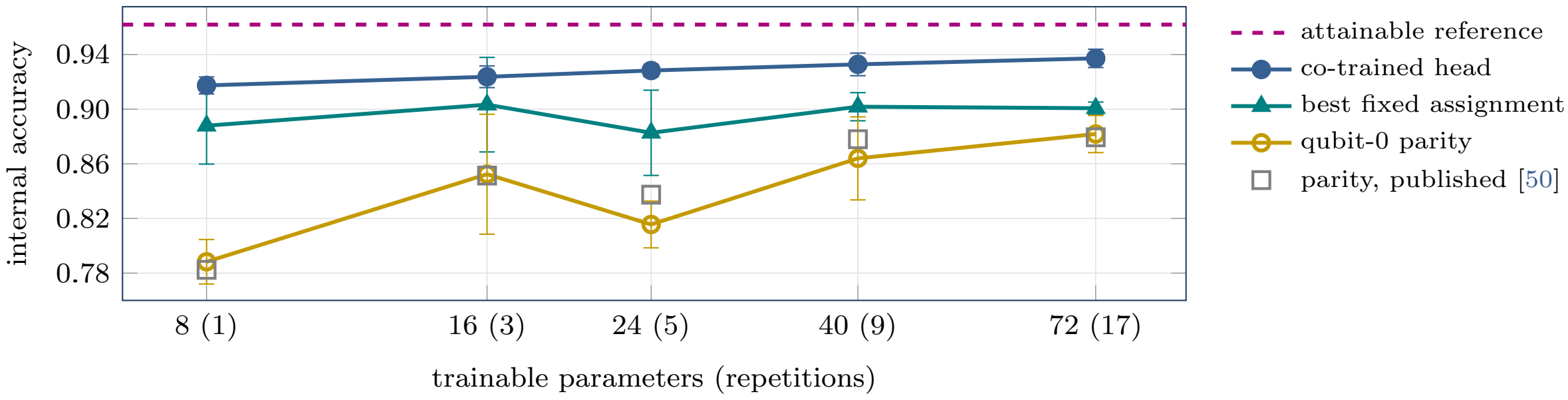

\noindent\textbf{Depth and readout.}
At the protocol of the control study (E5; Figure~\ref{fig: Depth}), the co-trained head is above qubit-0 parity at every depth, by 12.9, 7.1, 11.3, 6.9 and 5.5 points at 1, 3, 5, 9 and 17 repetitions (paired $p \leq 0.033$).
With 8 parameters and a co-trained head the circuit reaches $0.917 \pm 0.006$, above the 72-parameter circuit read by parity ($0.882 \pm 0.014$; $+3.6$ points, 95\% CI $+1.6$ to $+5.5$, $p = 0.007$).
With the head, accuracy rises by 1.1 points from 1 to 5 repetitions and by a further 0.9 points to 17, while the training objective falls from 1 to 17 repetitions under both readouts (parity 0.457 to 0.365, head 0.245 to 0.196).
The plateau that the control study observed over depth is therefore a plateau of the parity readout: the circuit's states hold more than parity extracts at every depth, and adding depth recovers only part of it.
% TONE-CHECK: Novel recall does not differ between the readouts at any depth ($p \geq 0.058$) and shows no monotone trend in depth (Spearman $\rho = 0.15$ for parity and $-0.12$ for the head).
Novel recall shows no significant difference between the readouts at any depth ($p \geq 0.058$) and no monotone trend in depth (Spearman $\rho = 0.15$ for parity and $-0.12$ for the head).

%--------------------------------------------------------------------------------------------------%
\subsection{What Do the Encodings Cost?}
%	\subsection{RQ3: What Do the Encodings Cost?}
%--------------------------------------------------------------------------------------------------%
\label{subsec: RQ3}
%--------------------------------------------------------------------------------------------------%

\begin{figure}[H]
	%----------------------------------------------------------------------------------------------%
	\centering
	%----------------------------------------------------------------------------------------------%
	\begin{tikzpicture}
		%------------------------------------------------------------------------------------------%
		\begin{axis}
			[
				name = ResourcesA,
				width = 0.360\linewidth, height = 4.400 cm,
				grid = major, grid style = { WordIceBlue, line width = 0.300 pt },
				axis line style = { WordBlueDarker50 }, tick label style = { font = \footnotesize },
				label style = { font = \footnotesize }, title style = { font = \footnotesize\bfseries },
				every axis plot/.append style = { line width = 1.100 pt, mark size = 2.000 pt },
				legend style = { font = \scriptsize, draw = none, fill = white, fill opacity = 0.85, text opacity = 1, cells = { anchor = west } },
				title = {(a) Two-qubit gates}, xlabel = { principal components $N$ },
				xmode = log, log basis x = 2, xmin = 3.3, xmax = 19.5, xtick = { 4, 8, 16 }, xticklabels = { 4, 8, 16 },
				ymin = 0, ymax = 60, ytick = { 0, 20, 40, 60 },
				legend to name = ResourcesLegend, legend columns = 3,
			]
			\addplot [ WordBlueDarker25, mark = * ] coordinates { (4, 2) (8, 18) (16, 44) };
			\addlegendentry{amplitude, all-to-all}
			\addplot [ WordBlueDarker25, dashed, mark = o ] coordinates { (4, 2) (8, 24) (16, 54) };
			\addlegendentry{amplitude, linear chain}
			\addplot [ MagentaDark, mark = square* ] coordinates { (4, 6) (8, 14) (16, 30) };
			\addlegendentry{angle, either connectivity}
		\end{axis}
		%------------------------------------------------------------------------------------------%
		\begin{axis}
			[
				name = ResourcesB,
				at = { ($ (ResourcesA.east) + (1.250 cm, 0) $) }, anchor = west,
				width = 0.360\linewidth, height = 4.400 cm,
				grid = major, grid style = { WordIceBlue, line width = 0.300 pt },
				axis line style = { WordBlueDarker50 }, tick label style = { font = \footnotesize },
				label style = { font = \footnotesize }, title style = { font = \footnotesize\bfseries },
				every axis plot/.append style = { line width = 1.100 pt, mark size = 2.000 pt },
				title = {(b) Depth}, xlabel = { principal components $N$ },
				xmode = log, log basis x = 2, xmin = 3.3, xmax = 19.5, xtick = { 4, 8, 16 }, xticklabels = { 4, 8, 16 },
				ymin = 0, ymax = 70, ytick = { 0, 20, 40, 60 },
			]
			\addplot [ WordBlueDarker25, mark = * ] coordinates { (4, 5) (8, 30) (16, 57) };
			\addplot [ WordBlueDarker25, dashed, mark = o ] coordinates { (4, 5) (8, 35) (16, 64) };
			\addplot [ MagentaDark, mark = square* ] coordinates { (4, 8) (8, 12) (16, 20) };
		\end{axis}
		%------------------------------------------------------------------------------------------%
		\begin{axis}
			[
				at = { ($ (ResourcesB.east) + (1.250 cm, 0) $) }, anchor = west,
				width = 0.360\linewidth, height = 4.400 cm,
				grid = major, grid style = { WordIceBlue, line width = 0.300 pt },
				axis line style = { WordBlueDarker50 }, tick label style = { font = \footnotesize },
				label style = { font = \footnotesize }, title style = { font = \footnotesize\bfseries },
				every axis plot/.append style = { line width = 1.100 pt, mark size = 2.000 pt },
				legend style = { font = \scriptsize, draw = none, fill = white, fill opacity = 0.85, text opacity = 1, cells = { anchor = west } },
				title = {(c) Finite shots, angle}, xlabel = { shots },
				xmode = log, xmin = 50, xmax = 200000, xtick = { 100, 1000, 10000, 100000 },
				ymin = -0.200, ymax = 1.650, ytick = { 0, 0.50, 1.00, 1.50 },
				yticklabel style = { /pgf/number format/fixed, /pgf/number format/precision = 2, /pgf/number format/zerofill },
				ylabel = { deficit (points) }, ylabel style = { yshift = -4pt },
				legend pos = north east, legend columns = 2,
				legend style = { font = \tiny, /tikz/every even column/.append style = { column sep = 2pt } },
			]
			\addplot [ WordGoldAccent1Darker25, mark = o ] coordinates { (100, 1.00) (1000, 0.52) (10000, 0.20) (100000, 0.12) };
			\addlegendentry{parity, $n = 8$}
			\addplot [ WordGoldAccent1Darker25, dashed, mark = triangle ] coordinates { (100, 0.43) (1000, 0.06) (10000, 0.06) (100000, -0.03) };
			\addlegendentry{parity, $n = 16$}
			\addplot [ WordBlueDarker25, mark = * ] coordinates { (1000, 0.24) (10000, 0.12) };
			\addlegendentry{head, $n = 8$}
			\addplot [ WordBlueDarker25, dashed, mark = triangle* ] coordinates { (1000, 0.12) (10000, -0.04) };
			\addlegendentry{head, $n = 16$}
		\end{axis}
		%------------------------------------------------------------------------------------------%
	\end{tikzpicture}
	%----------------------------------------------------------------------------------------------%
	\\[0.150 cm]
	\pgfplotslegendfromname{ResourcesLegend}
	%----------------------------------------------------------------------------------------------%
	\caption{Cost of the encodings.
	(a) Two-qubit gates and (b) depth of the matched circuits (E4 layout, $3N$ parameters) after transpilation to $\{\mathrm{CX}, U\}$ with Qiskit 2.3.0 at optimisation level 3 (E7; amplitude state preparation: median over 25 records).
	(c) Accuracy lost by the seed-0 angle-encoded circuits when their readout is estimated from finite shots (E6): exact minus sampled accuracy, parity with global parity over 20 draws on 2,000 records, co-trained head over 5 draws on 500 records.
	%	Values in Tables~S6 and S8a.
	}
	\label{fig: Resources}
	%----------------------------------------------------------------------------------------------%
\end{figure}
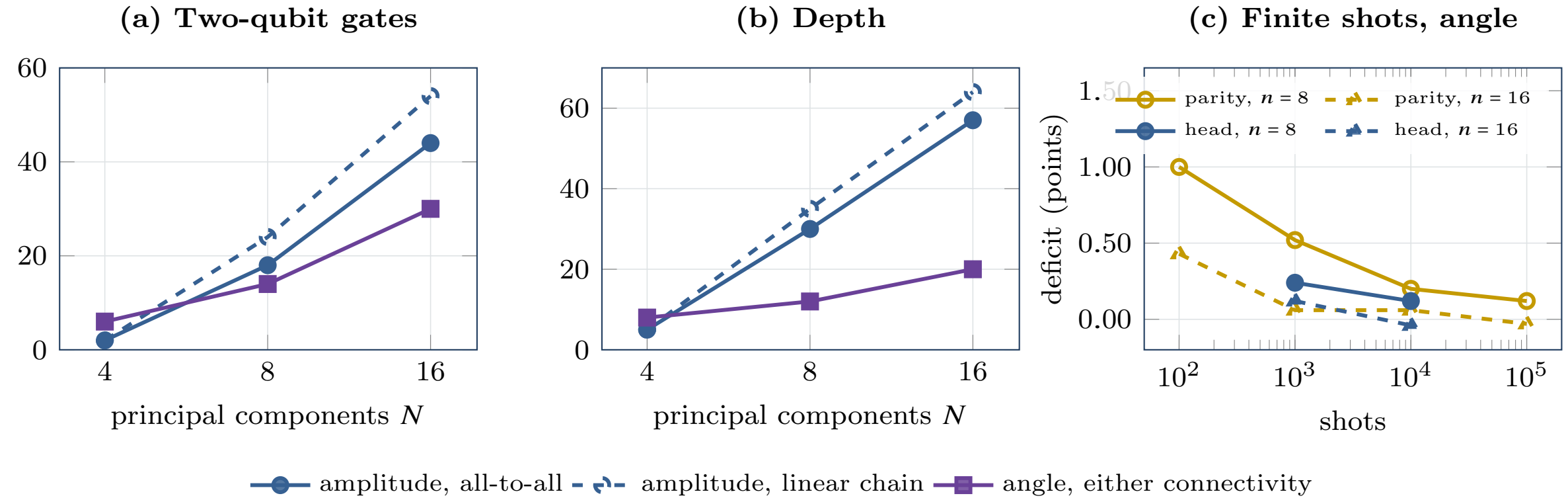

After transpilation (Figure~\ref{fig: Resources}a,b), amplitude encoding needs $\log_{2} N$ qubits but a state preparation whose two-qubit gate count equals $2^{n} - n - 1$ on $n$ qubits (1, 4, 11, 26, 57, 120 and 247 for $n = 2$ to 8), so its circuits are deeper: at matched parameter count the amplitude-encoded circuits have 18 and 44 two-qubit gates at depth 30 and 57 at $N = 8$ and 16, against 14 and 30 at depth 12 and 20 for the angle-encoded circuits, which need $N$ qubits but no two-qubit gate to encode. On a linear chain of qubits the amplitude circuit grows to 54 two-qubit gates, while the angle circuit, whose entangling gates already follow a chain, does not change.
At $N = 4$ the amplitude circuit is the cheaper of the two.

Read with a finite number of shots (E6; Figure~\ref{fig: Resources}c), the parity readout of the seed-0 angle-encoded circuits approaches its exact accuracy quickly: at $n = 8$ the deficit is 1.00, 0.52, 0.20 and 0.12 points at $10^{2}$, $10^{3}$, $10^{4}$ and $10^{5}$ shots, and at $n = 16$ it is below 0.5 points from $10^{2}$ shots on. The co-trained head, with all its features estimated from the same shots, loses 0.24 ($n = 8$) and 0.12 points ($n = 16$) at $10^{3}$ shots and at most 0.12 points at $10^{4}$. The cost of training on a device is set by the protocol rather than by the readout: evaluations $\times$ records $\times$ shots gives $3.0 \times 10^{8}$ circuit executions for the protocol of the control study and $8.0 \times 10^{8}$ for that of E4 at $10^{3}$ shots.

%--------------------------------------------------------------------------------------------------%
%--------------------------------------------------------------------------------------------------%

%--------------------------------------------------------------------------------------------------%
%                                                                                                  %
%			W h e r e   D o e s   A                                                                %
%			V a r i a t i o n a l   Q u a n t u m   C l a s s i f i e r                            %
%			L o s e   A c c u r a c y   ?                                                          %
%                                                                                                  %
%--------------------------------------------------------------------------------------------------%
%                                                                                                  %
%			S e c t i o n   5                                                                      %
%                                                                                                  %
%--------------------------------------------------------------------------------------------------%
%                                                                                                  %
%			D i s c u s s i o n                                                                    %
%                                                                                                  %
%--------------------------------------------------------------------------------------------------%
\section{Discussion}
%--------------------------------------------------------------------------------------------------%
\label{sec: Discussion}
\subsection{What the Attainable Reference Shows}
%--------------------------------------------------------------------------------------------------%
\label{subsec: Discussion Reference}
%--------------------------------------------------------------------------------------------------%

The reference separates two questions that a single accuracy figure mixes: how much the encoded states allow, and how much of it a trained circuit takes. On the first, the two encodings are within one point at 16 components although angle encoding retains five to six times the pairwise signal; a weak AUC does not preclude high accuracy, and on this benchmark a linear model on 16 components already reaches 0.951.
%--------------------------------------------------------------------------------------------------%
% TONE-CHECK: On the second they differ: with a co-trained linear readout an angle-encoded circuit reaches its reference at every size, an amplitude-encoded one stops 0.7 to 1.3 points short, and with parity both stop further away.
%--------------------------------------------------------------------------------------------------%
On the second they differ: with a co-trained linear readout an angle-encoded circuit comes within 0.3 points of its reference ($C = 1$) or exceeds it at every size, an amplitude-encoded one stops 0.7 to 1.3 points short, and with parity both stop further away. The readout is the largest measured loss in both encodings; what remains after co-training, at most 1.3 points, is the joint share of the ansatz, the optimizer budget and the finite training set. At 16 components the trained advantage of angle encoding therefore comes from how much of its reference each circuit realises, not from the geometry. The design attributes it to the encoding together with what it implies (16 qubits instead of 4, a head of 137 features instead of 11, a 65,536-dimensional state space instead of 16); readout dimension alone does not explain it, since with a single observable (global parity) the gap is larger.
%--------------------------------------------------------------------------------------------------%
% REFS-CUT: The reference is not an upper bound: the non-linear MLP head exceeds its $C = 1$ value by 1.3 points and the co-trained angle circuits by up to 0.6 points, and it is distinct from the Helstrom limit for single-copy discrimination \cite{helstrom1969}.
%--------------------------------------------------------------------------------------------------%
The reference is not an upper bound: the non-linear MLP head exceeds its $C = 1$ value by 1.3 points and the co-trained angle circuits by up to 0.6 points.

\noindent\textbf{Quantum and classical models.}
The kernels of both encodings have closed forms, so the attainable reference is computed classically: an SVM with a homogeneous quadratic kernel for amplitude encoding and, at this bandwidth, practically a Gaussian-kernel SVM for angle encoding, which is what bandwidth-tuned quantum kernels are known to approach \cite{Kubler2021, Canatar2022, Shaydulin2022, Slattery2023, Florezablan2025}. Every circuit in this study therefore has a classical yardstick whose cost grows only linearly with the number of components. Measured against it and against logistic regression on the same inputs, the angle-encoded circuits with a co-trained head are 0.4 to 1.0 points more accurate than logistic regression and 0.5 to 1.9 points less accurate than the SVM with $C = 100$, and every circuit read by parity alone is on average 2.0 to 11.9 points less accurate than logistic regression. Out of distribution, the novel-signature recall of the amplitude-encoded circuits is 9.0 and 10.4 points above that of logistic regression at $N = 4$ and 16 ($p \leq 0.032$); the amplitude reference recalls 9.7 points more than the circuits at $N = 4$ ($p < 0.001$) and does not differ from them significantly at $N = 8$ and 16 ($p \geq 0.08$).
%--------------------------------------------------------------------------------------------------%
% REFS-CUT: On this benchmark the trained circuits therefore perform at the level of the classical models that share their feature map, in line with benchmark studies \cite{Bowles2024, alvarezestevez2025, Schnabel2025} and with work on classical surrogates \cite{Schreiber2023, Sweke2025} and generalisation \cite{caro2022, gilfuster2024}.
%--------------------------------------------------------------------------------------------------%
On this benchmark the trained circuits therefore perform at the level of the classical models that share their feature map, in line with benchmark studies \cite{Bowles2024, Alvarezestevez2025, Schnabel2025} and with work on classical surrogates \cite{Schreiber2023, Sweke2025} and generalization \cite{GilFuster2024}.
%--------------------------------------------------------------------------------------------------%
% REFS-CUT: The measurements also indicate where a difference can arise: a quantum kernel that a classical computer evaluates efficiently cannot by itself provide a learning advantage \cite{HuangHY2021, Kubler2021}, so the protocol is most informative for encodings whose kernels have no efficient classical form, such as the feature maps proposed for that purpose \cite{Havlicek2019}; for them the reference is estimated on quantum hardware \cite{havlicek2019, hubregtsen2022}.
%--------------------------------------------------------------------------------------------------%
The measurements also indicate where a difference can arise: a quantum kernel that a classical computer evaluates efficiently cannot by itself provide a learning advantage \cite{HuangHY2021, Kubler2021}, so the protocol is most informative for encodings whose kernels have no efficient classical form, such as the feature maps proposed for that purpose \cite{Havlicek2019}; for them the reference is estimated on quantum hardware \cite{Havlicek2019}.

%--------------------------------------------------------------------------------------------------%
%                                                                                                  %
%			W h e r e   D o e s   A                                                                %
%			V a r i a t i o n a l   Q u a n t u m   C l a s s i f i e r                            %
%			L o s e   A c c u r a c y   ?                                                          %
%                                                                                                  %
%--------------------------------------------------------------------------------------------------%
%                                                                                                  %
%			S u b s e c t i o n   5 . 2                                                            %
%                                                                                                  %
%--------------------------------------------------------------------------------------------------%
%                                                                                                  %
%			R e a d i n g   T h e                                                                  %
%			T w o   S t u d i e s   T o g e t h e r                                                %
%                                                                                                  %
%--------------------------------------------------------------------------------------------------%
\subsection{Reading the Two Studies Together}
%--------------------------------------------------------------------------------------------------%
\label{subsec: Two Studies}
%--------------------------------------------------------------------------------------------------%

The new measurements bear on each statement of the control study \cite{Thomos2026}. The results and the re-run agree within their seed-to-seed dispersion (Section~\ref{subsec: Anchor}); the preprocessing of Section~\ref{subsec: Data} adds about two points to a linear classifier.

\noindent\textbf{Angle encoding} as a direction is supported, for a reason other than the one the AUC suggested: under the matched protocol it gives more accurate trained circuits at every size, by more than its geometric advantage at 16 components.

\noindent\textbf{The readout}, which the control study pointed to but could not separate from the encoding, is the largest measured loss.

\noindent\textbf{The depth plateau} is a readout plateau: a co-trained head lifts an 8-parameter circuit above the 72-parameter parity circuit, and once the readout is trained depth adds about two points.

\noindent\textbf{Norm degeneracy} is not the operative mechanism at this scale, although five seeds cannot exclude a gain of up to about five points.

The higher recall on unseen attack types, observed there without a mechanism, holds for amplitude-encoded circuits with every linear readout and at every depth tested, and belongs to the encoding, since the amplitude reference shows it too. The numbers are consistent with locality: angle-encoded models, whose kernel is close to a Gaussian, recall about as many novel records as a nearest-neighbor rule on the same fidelity, whereas the quadratic amplitude kernel, which depends only on the direction of a record, generalities further from the training attacks.
\subsection{Consequences for Hardware}
%--------------------------------------------------------------------------------------------------%
\label{subsec: Hardware}
%--------------------------------------------------------------------------------------------------%

%--------------------------------------------------------------------------------------------------%
% REFS-CUT: The encoding sets the qubit count and the state-preparation cost, whose gate count follows $2^{n} - n - 1$ exactly, consistent with the exponential worst-case scaling of exact preparation \cite{Mottonen2005, Shende2006, Plesch2011, iten2016, Sun2023}.
%--------------------------------------------------------------------------------------------------%
The encoding sets the qubit count and the state-preparation cost, whose gate count follows $2^{n} - n - 1$ exactly, consistent with the exponential worst-case scaling of exact preparation \cite{Mottonen2005, Shende2006, Plesch2011, Sun2023}. At matched parameters the angle-encoded circuits at $N = 8$ and 16 need fewer two-qubit gates and less than half the depth, and would accumulate 22\% and 32\% fewer two-qubit-gate errors at a fixed error per gate (14 against 18 and 30 against 44 two-qubit gates); amplitude encoding is favoured by qubit count (4 against 16) and at $N = 4$.
%--------------------------------------------------------------------------------------------------%
% REFS-CUT: Which matters more depends on the device and its noise \cite{preskill2018, cai2023}, which can flatten the training landscape \cite{Wang2021} and was not simulated.
Which matters more depends on the device and its noise, which can flatten the training landscape %--------------------------------------------------------------------------------------------------%
\cite{Wang2021} and was not simulated.
The readout result is favourable: a linear head over Pauli-$Z$ strings is computed from the same computational-basis samples as parity and is as tolerant to finite shots.
%--------------------------------------------------------------------------------------------------%
% REFS-CUT: Training on a device is the larger cost, at $10^{8}$ to $10^{9}$ circuit executions, and a stochastic objective changes the optimisation problem itself \cite{sweke2020}.
%--------------------------------------------------------------------------------------------------%
Training on a device is the larger cost, at $10^{8}$ to $10^{9}$ circuit executions.

%--------------------------------------------------------------------------------------------------%
%                                                                                                  %
%			W h e r e   D o e s   A                                                                %
%			V a r i a t i o n a l   Q u a n t u m   C l a s s i f i e r                            %
%			L o s e   A c c u r a c y   ?                                                          %
%                                                                                                  %
%--------------------------------------------------------------------------------------------------%
%                                                                                                  %
%			S u b s e c t i o n   5 . 4                                                            %
%                                                                                                  %
%--------------------------------------------------------------------------------------------------%
%                                                                                                  %
%			R e a d i n g   T h e                                                                  %
%			R e f e r e n c e   S h o w s                                                          %
%                                                                                                  %
%--------------------------------------------------------------------------------------------------%
\subsection{Limitations}
%--------------------------------------------------------------------------------------------------%
\label{subsec: Limitations}
%--------------------------------------------------------------------------------------------------%

The 16-qubit angle circuit was trained for one seed and not with qubit-0 parity, so comparisons at $N = 16$ have no intervals.
The matched comparison cannot separate the encoding from its consequences (qubit count, state-space dimension, head size). Only one angle scale ($s = 0.25$, the kernel bandwidth) was studied, and the out-of-distribution result may depend on it.
%--------------------------------------------------------------------------------------------------%
% REFS-CUT: All results are noiseless, shots were applied at inference only, and trainability at larger qubit counts was not probed \cite{McClean2018, larocca2025}.
%--------------------------------------------------------------------------------------------------%
All results are noiseless, shots were applied at inference only, and trainability at larger qubit counts was not probed \cite{McClean2018}.
%--------------------------------------------------------------------------------------------------%
% REFS-CUT: One legacy benchmark with binary labels was used; newer datasets \cite{moustafa2015, sharafaldin2018} and multi-class imbalance \cite{luque2019} were not addressed.
%--------------------------------------------------------------------------------------------------%
One legacy benchmark with binary labels was used; newer datasets and multi-class imbalance were not addressed. Five seeds support description, not equivalence.
%--------------------------------------------------------------------------------------------------%

%--------------------------------------------------------------------------------------------------%
%                                                                                                  %
%	C o n c l u s i o n s                                                                          %
%                                                                                                  %
%--------------------------------------------------------------------------------------------------%
\section{Conclusions} \label{sec: Conclusions}

Where do variational classifiers lose accuracy? In this benchmark, the loss occurs primarily at the readout—the final step converting measurement outcomes into a class label. This limitation applies to both encodings and accounts for the depth plateau observed in our earlier classifier, as anticipated by the prior study. A co-trained readout uses the same measurements as the parity observable to recover most of the lost accuracy, making it the primary component of a variational classifier worthy of optimization.

The optimal encoding choice depends on the specific task. Angle encoding yields a more accurate detector for traffic similar to the training data and, for dimensions of eight components or higher, requires shallower circuits. Conversely, amplitude encoding uses fewer qubits and detects a higher proportion of attack types absent from the training data. On identical inputs, the quantum circuits perform on par with classical models sharing the same feature map.

Whether quantum feature maps can achieve superior performance remains an open question for encodings whose kernels cannot be efficiently evaluated by classical computers \cite{Havlicek2019, HuangHY2021}. Below, we summarize the key conclusions from our extensive experimental analysis.

%--------------------------------------------------------------------------------------------------%
\begin{MyInfoBox}[Synopsis]				%	Title: Synopsis
	%----------------------------------------------------------------------------------------------%
	\begin{enumerate}
		%------------------------------------------------------------------------------------------%
		[ left = 0.500 cm, labelsep = 0.500 cm, start = 1 ]
		\renewcommand \labelenumi { (\textbf{C}$_{ \theenumi }$) }
		%------------------------------------------------------------------------------------------%
		\item
		\textbf{Encoding.}
		Angle encoding retains five to six times the pairwise class signal of amplitude encoding at $16$ components; however, the attainable reference accuracies of the two encodings differ by at most one percentage point at that scale, compared to a $3.5$ to $3.7$ percentage point difference at four components.
		%------------------------------------------------------------------------------------------%
		\item
		\textbf{Classifier.}
		With a co-trained linear readout, angle-encoded circuits approach their reference values ($C = 1$: $0.930$, $0.960$, and $0.965$ versus $0.933$, $0.956$, and $0.959$) and outperform amplitude-encoded circuits by $1.6$ to $4.5$ percentage points under a given protocol. Global parity leaves $2.6$ to $7.4$ percentage points unused, and qubit-$0$ parity leaves $11.4$ percentage points unused on the published circuit, indicating that the readout is the primary source of loss. Thus, the depth plateau in the control study is fundamentally a readout plateau.
		%------------------------------------------------------------------------------------------%
		\item
		\textbf{Out of distribution.}
		Angle-encoded classifiers recall $0.30$ to $0.45$ of unseen attack types, compared to $0.55$ to $0.61$ for amplitude-encoded ones. This performance trade-off is inherent to the encodings and aligns with the locality of the angle kernel.
		%------------------------------------------------------------------------------------------%
		\item
		\textbf{Quantum and classical models.}
		The best circuits achieve $0.4$ to $1.0$ percentage points higher accuracy than logistic regression on identical inputs, but are $0.5$ to $1.9$ percentage points less accurate than an SVM with $C = 100$ using its own kernel. Out of distribution, amplitude-encoded circuits recall $9.0$ and $10.4$ percentage points more unseen attacks than logistic regression at $N = 4$ and $N = 16$, respectively.
		%------------------------------------------------------------------------------------------%
		\item
		\textbf{Resources.}
		At $N = 16$ and matched parameters, the angle-encoded circuit requires $16$ qubits, $30$ two-qubit gates, and a depth of $20$, whereas the amplitude-encoded circuit requires $4$ qubits, $44$ two-qubit gates, and a depth of $57$.
		%------------------------------------------------------------------------------------------%
	\end{enumerate}
	%----------------------------------------------------------------------------------------------%
\end{MyInfoBox}
%--------------------------------------------------------------------------------------------------%

It is expedient to compute the encoding's reference performance prior to training subsequent circuits. This estimates the upper bound of accuracy permitted by the encoding, while the distance between the trained circuit and this reference quantifies the performance margin left unutilized by the circuit architecture.

%--------------------------------------------------------------------------------------------------%
%                                                                                                  %
%			W h e r e   D o e s   A                                                                %
%			V a r i a t i o n a l   Q u a n t u m   C l a s s i f i e r                            %
%			L o s e   A c c u r a c y   ?                                                          %
%                                                                                                  %
%--------------------------------------------------------------------------------------------------%
%                                                                                                  %
%			S u b s e c t i o n                                                                    %
%                                                                                                  %
%--------------------------------------------------------------------------------------------------%
%                                                                                                  %
%			D a t a   A v a i l a b i l i t y                                                      %
%                                                                                                  %
%--------------------------------------------------------------------------------------------------%
\subsection*{Data Availability}
%--------------------------------------------------------------------------------------------------%
\label{subsec: Data Availability}
%--------------------------------------------------------------------------------------------------%

The code, result files and analysis scripts of this study are available at \url{https://github.com/Angeloth1/vqcAngle}, and those of the control study at \url{https://github.com/Angeloth1/Quantum-IDS-NSLKDD}.
Every number in the Results is produced by these scripts from the stored result files.
% The works counted in Figure~\ref{fig: Literature} are listed in Dataset S1, from which the literature-count script in the same repository recomputes every value of the figure.
NSL-KDD is publicly available \cite{Tavallaee2009}.

\bibliographystyle{ieeetr}
%                                                                                                  %
\bibliography{WVQCLA}
%                                                                                                  %
%--------------------------------------------------------------------------------------------------%
%--------------------------------------------------------------------------------------------------%
%--------------------------------------------------------------------------------------------------%

%--------------------------------------------------------------------------------------------------%
%--------------------------------------------------------------------------------------------------%
%--------------------------------------------------------------------------------------------------%
%                                                                                                  %
%			T h e   A r t i c l e   E n d s   H e r e                                              %
%                                                                                                  %
%--------------------------------------------------------------------------------------------------%
%--------------------------------------------------------------------------------------------------%
%--------------------------------------------------------------------------------------------------%
%                                                                                                  %
\end{document}